\documentclass[floatfix,aps,prl,amsmath,amssymb,twocolumn,showpacs,superscriptaddress,secnumarabic]{revtex4-2}
\usepackage{verbatim}
\usepackage{graphicx}  
\usepackage{dcolumn}   
\usepackage{bm}   
\usepackage{times}
\usepackage{physics}
\usepackage{siunitx}
\usepackage[utf8]{inputenc}
\usepackage[T1]{fontenc}
\usepackage[colorlinks, citecolor=blue]{hyperref}
\usepackage[utf8]{inputenc}
\usepackage[english]{babel}
\usepackage{multirow}
\usepackage{amsmath}
\usepackage{amssymb}
\usepackage{mathrsfs}
\usepackage{textcomp}
\usepackage{tabularx}
\usepackage{color}
\usepackage{array}
\usepackage{float}
\usepackage[dvipsnames]{xcolor}
\definecolor{darkblue}{rgb}{0.1,0.2,0.6}
\definecolor{darkred}{rgb}{0.8,0.1,0.2}

\begin{document}

\title{Probing the topological protection of edge states in multilayer tungsten ditelluride with the superconducting proximity effect}
\author{X. Ballu$^{1,5}$, Z. Dou$^{1,4}$, L. Bugaud$^1$, R. Delagrange$^1$, A. Bernard$^1$, Ratnadwip Singha$^{2,3}$, L. M. Schoop$^2$, R. J.  Cava$^2$,  R. Deblock$^1$, Sophie Gu\'eron$^1$, H. Bouchiat$^1$ and M. Ferrier$^1$}
\affiliation{$^1$Université Paris-Saclay, CNRS, Laboratoire de Physique des Solides, 91405, Orsay, France.}
\affiliation{$^2$Department of Chemistry, Princeton University, Princeton, NJ 08544, USA.}
\affiliation{$^3$Current address: Department of Physics, Indian Institute of Technology Guwahati, Assam 781039, India.}
\affiliation{$^4$ Current address: Beijing National Laboratory for Condensed Matter Physics and Institute of Physics, Chinese Academy of Sciences, Beijing 100190, China}
\affiliation{$^5$ Current address: Laboratoire National de M\'etrologie et d'Essais, 29 avenue Roger Hennequin, 78197 Trappes, France}
 
\begin{abstract}
	
The topology of WTe$_2$, a transition metal dichalcogenide with large spin-orbit interactions, is thought to combine type II Weyl semimetal and second-order topological insulator (SOTI) character. The SOTI character should endow WTe$_2$ multilayer crystals with topologically protected helical states at its hinges, and, indeed, 1D states have been detected thanks to Josephson interferometry. However, the immunity to backscattering conferred to those states by their helical nature has so far not been tested.
To probe the topological protection of WTe$_2$ edge states, we have fabricated Superconducting Quantum Interference Devices (SQUIDs) in which the supercurrent through a junction on the crystal edge interferes with the supercurrent through a junction in the bulk of the crystal. We find behaviors ranging from a Symmetric SQUID to asymmetric SQUID patterns, including one in which the modulation by magnetic field reveals a sawtooth-like supercurrent versus phase relation for the edge junction, demonstrating that the supercurrent at the edge is carried by ballistic  channels over 600 nm, a tell-tale sign of the SOTI character of multilayer WTe$_2$. 
\end{abstract}

\maketitle

\section{INTRODUCTION: DETECTING HELICAL STATES IN WTe$_2$}
Three dimensional (3D) Second Order Topological Insulators (SOTIs) are a relatively newly discovered phase of matter, which possesses the interesting property of hosting topologically protected one-dimensional states. These states stem from a double band inversion in the material, and spatial symmetries in the bulk that are broken at the crystal's surfaces, turning them insulating, with surface gaps that can alternate between being inverted and not. The topological one-dimensional states are expected at the hinges between such surfaces. One of the most interesting properties of these states is their ballistic conduction over long distances even in the presence of disorder. This is due to the protection offered by the spin-momentum locking characteristic of the helical nature of topological quantum Spin Hall states, and contrasts with the Anderson localisation by disorder of 1D states that have no such topological protection. Their identification is not straightforward, however, in part because more numerous non-topological states often coexist with the 1D helical states. 

 \begin{figure}[h!]
	\centering
	\includegraphics[width=8 cm]{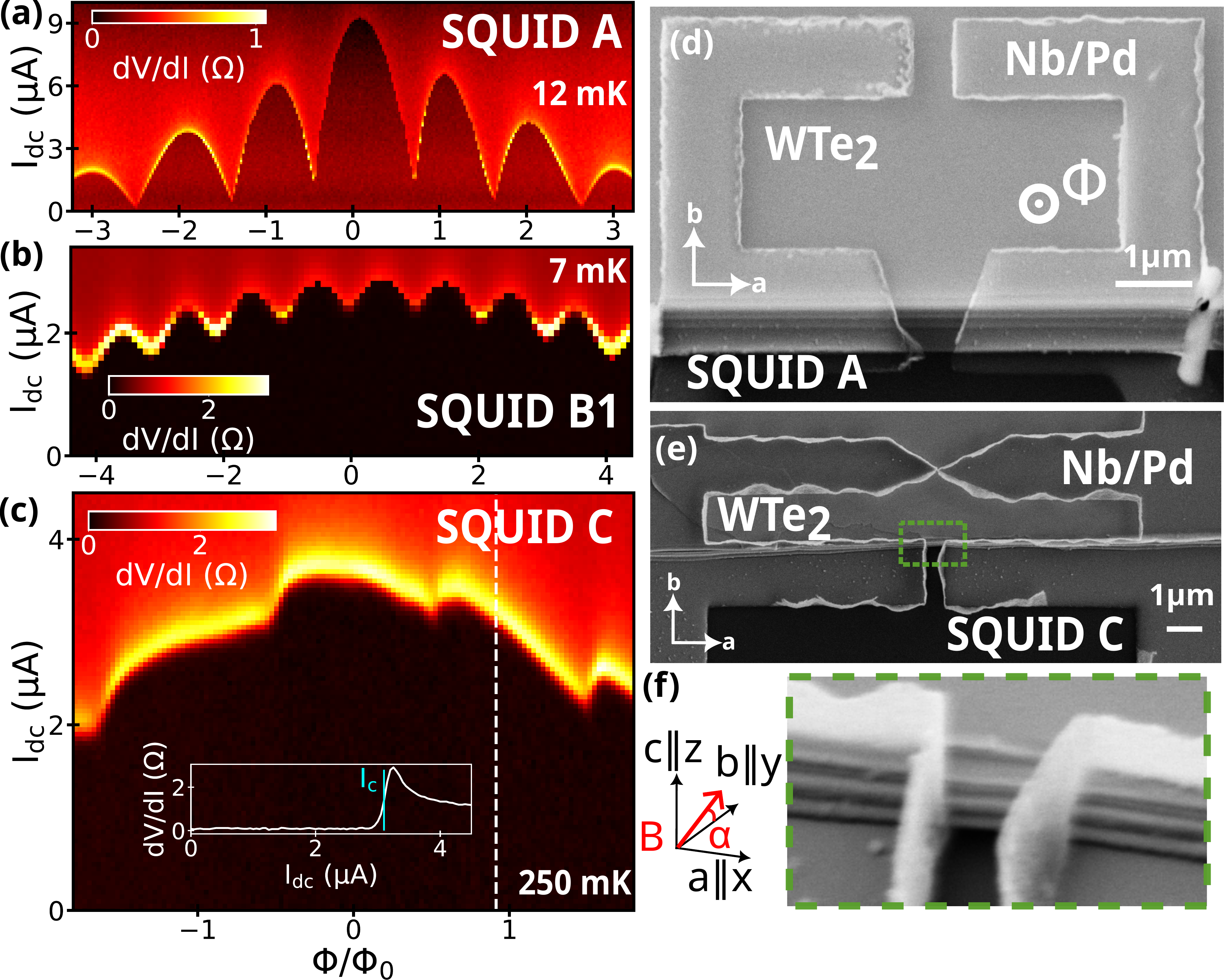}
	\caption{WTe$_2$-based SQUIDs A,B1,C and their interference patterns. (a)-(c): Colour-coded differential resistance $ dV/dI$ as function of the dc current $I_{DC}$ through the bulk and edge junctions in parallel, and the magnetic flux $\Phi$ through the SQUID loop, in units of the flux quantum $\Phi_0=h/2e$. The critical current $I_c$ corresponds to the boundary between the zero (black) and finite (red) resistance regions. Dilution refrigerator temperatures are indicated in the graphs. (a) SQUID A displays a practically full modulation of $I_c$, indicative of a symmetric SQUID. (b) SQUID B1 displays a small sinusoidal-like modulation of $I_c$. (c)  SQUID C's $I_c$ (defined in inset linecut at fixed $\Phi$, indicated by white dashed line) has a sawtooth-like modulation, atop a smooth background, a tell-tale sign of ballistic transport. The leads resistance has been subtracted. (d),(e) Scanning Electron Micrograph of SQUIDs $A$ and $C$, with (f) a zoom of SQUID $C$'s edge junction, and a sketch of the magnetic field $B$'s orientation relative to the crystal axes.  The magnetic field $B$ is in the $(b,c)$ plane, at a small angle $\alpha $ with the $b$-axis. Thus $\Phi=B_zS\simeq \alpha BS$, with $S\simeq 40 \mathrm{~\mu m^2}$ the SQUID loop area, which includes focusing effects (see appendix). $\alpha\simeq 0.02 \mathrm{~rad}$ is deduced from the oscillation period $B_0$ via $\alpha=\Phi_0/B_0S$.  
 }
	\label{fig:Fig_1}
\end{figure}

Among the proposed 3D SOTIs, $\mathrm{WTe_2}$ is a transition-metal dichalcogenide (TMD) whose monolayer has been shown to possess a 2D topological insulator (TI) character, with a large insulating bulk gap (in the tens of meV range), and quantum spin Hall edge conduction \cite{Fei2016,wu2018a,peng_observation_2017,tang2017,shi2019,Maximenko2022}. The 2D TI character stems from inversion symmetry combined with band inversion at the gamma point. The topology of multilayer WTe$_2$ is less clear. Its $T_d$ phase (symmetry group $Pmn2_1$) is characterized by a double band inversion at the gamma point. Combined with the screw symmetry along the $c$-axis, this can lead to a bulk 3D topological state in which helical modes can exist at surface domain walls and hinges. The precise configuration of these helical modes depends on a combination of surface band inversion and disorder. In practice, like in Bi$_4$Br$_4$ \cite{BiBr_Hsu2019,BiBr_Lin2024,BiBr_Tang2019,BiBr_Yoon2020}, helical states are likely to appear along the $a$ and $b$ directions on alternating terraces or step edges, see Fig. \ref{fig:Fig_1} and Fig. 30 
and \cite{wang2019a,song_natcom_2018,wieder_science2018,wieder_double_2016}. Observing these helical modes can however be made difficult by the (type II Weyl) semi-metallic nature of WTe$_2$: electron and hole pockets at the Fermi level provide a high density of trivial bulk carriers that may obscure the signatures of the fewer topological modes. Nevertheless, signatures of topological hinge states have been reported: Optical Kerr rotation microscopy detected spin polarization confined to hinges of $\mathrm{WTe_2}$  \cite{lee2023}. In mesoscopic samples with superconducting contacts, proximity-induced superconductivity enhances the relative contribution of topological states with respect to trivial states \cite{Murani2016}. The critical current (maximal induced supercurrent) dependence on magnetic field, also called Josephson interferometry, has been exploited to  reveal supercurrents flowing at the edges of a sample rather than in the bulk \cite{Li2014a,Hart2014,Li2019,kononov2020,choi2020}. 
In addition, the critical current's decay with field depends on the number and width of the supercurrent-carrying modes: nanometer-size 1D channels can carry supercurrent over micrometers up to fields of several Teslas, as was demonstated in bismuth nanowires \cite{Li2014a,Murani2016}. 
In  WTe$_2$, Josephson interferometry suggests that the supercurrent is carried preferentially along the $a$-direction crystal edges \cite{endres2022,choi2020}. However, Josephson interferometry cannot measure the degree of protection against backscattering, a key signature of helical transport. Instead, the measurement of a supercurrent-versus-phase relation (CPR) can provide such a signature, because its shape is exquisitely sensitive to the transport regime: CPRs of junctions with no backscattering display a jump at $\pi$ phase difference, in stark contrast to the smooth, sinusoidal-like CPRs of tunnel or diffusive junctions. A sawtooth-shaped CPR is the signature of a long ballistic Josephson junction and has been found in bismuth- and BiSb-based Josephson junctions \cite{Murani2016,Li2019}. Interestingly, CPRs with jumps at $\pi$ are robust to an imperfect NS interface only if there is topological protection \cite{murani_andreev_2017,cayao_andreev_2018,SM}, see appendix A. In addition, CPRs can probe the dynamics of Andreev states, providing a test of their helical character \cite{Bernard2023}.

\section{DETERMINATION OF THE SUPERCURRENT VERSUS PHASE RELATION USING ASYMMETRIC SQUIDs}

To determine the CPR, we fabricate Superconducting Quantum Interference Devices (SQUIDs) in which the critical current results from interference between one Josephson junction on an $a$-oriented crystal edge, and another Josephson junction in the bulk of the crystal's $(001)$ surface. This configuration is somewhat similar to \cite{endres2023}, except that we design devices with supercurrents small enough that distortions of the critical current by screening effects are negligible (see Appendix E). 
We find different behaviors, ranging from symmetric to asymmetric SQUID patterns. The interference pattern of one asymmetric SQUID, SQUID C, with the shortest edge junction,  reveals a sawtooth-shaped CPR. The temperature and field dependences of SQUID C's critical current, and notably its harmonics content, demonstrate that the supercurrent at the edge is carried quasi-ballistically over 600 nm, a tell-tale sign of the SOTI character of WTe$_2$. From the decay with field of the supercurrent, we determine that the edge states extend less than $2\ $nm in the $c$ direction, and $20\ $nm in the (a,b) plane. The main text focuses on this SQUID C, while in the appendix we present the other SQUIDs and offer explanations for the range of behaviors observed. 

The SQUIDs are formed by sputter-deposition of a (8 nm/80 nm)-thick Pd/Nb superconducting film onto 30- to 300-nm-thick WTe$_2$ flakes obtained by exfoliation in a glove box of high-quality crystals with a residual resistivity ratio exceeding 1000  \cite{Cava2014,SM}. The use of Pd favors good contacts to the WTe$_2$, thanks to the superconducting $PdTe_x$ compound which forms by interdiffusion of Pd in WTe$_2$ \cite{endres2022,ohtomo2022,jia_superconductivity_2024}. The SQUIDs are designed such that one junction, formed in the bulk of the crystal surface, acts as a reference junction with a higher critical current $I^{ref}_{c}$, while the other junction, on the crystal edge, is expected to have a smaller critical current $I^{edge}_{c}$, predominantly carried by the edge states, see Fig. \ref{fig:Fig_1}. For such an asymmetric SQUID in a magnetic field $B$ with $B_z$ projection perpendicular to the SQUID loop plane, the critical current reads \cite{DellaRocca2007}:
$$I_c(B)\simeq I^{ref}_{c}(B)+I_{0}(B)f(\varphi_0-2\pi\frac{B_zS}{\Phi_0}),$$ with $I^{edge}(B,\varphi)=I_0(B)f(\varphi)$ the edge junction's CPR, $S$ the area defined by the SQUID loop, $\Phi_0=h/2e$ the superconducting flux quantum and $\varphi_0$ the phase at which the reference junction's CPR is maximal. For such an asymmetric SQUID, $I_c$ is then practically a direct measurement of the edge junction's CPR. 
By contrast, if the SQUID is symmetric, $I^{edge}_c \simeq I^{bulk}_c$, a different, even-in-field interference pattern emerges, with full periodic cancellation of $I_c$ in the case of a sinusoidal CPR.

\begin{figure*}[htb]
	\centering
	\includegraphics[width=12 cm]{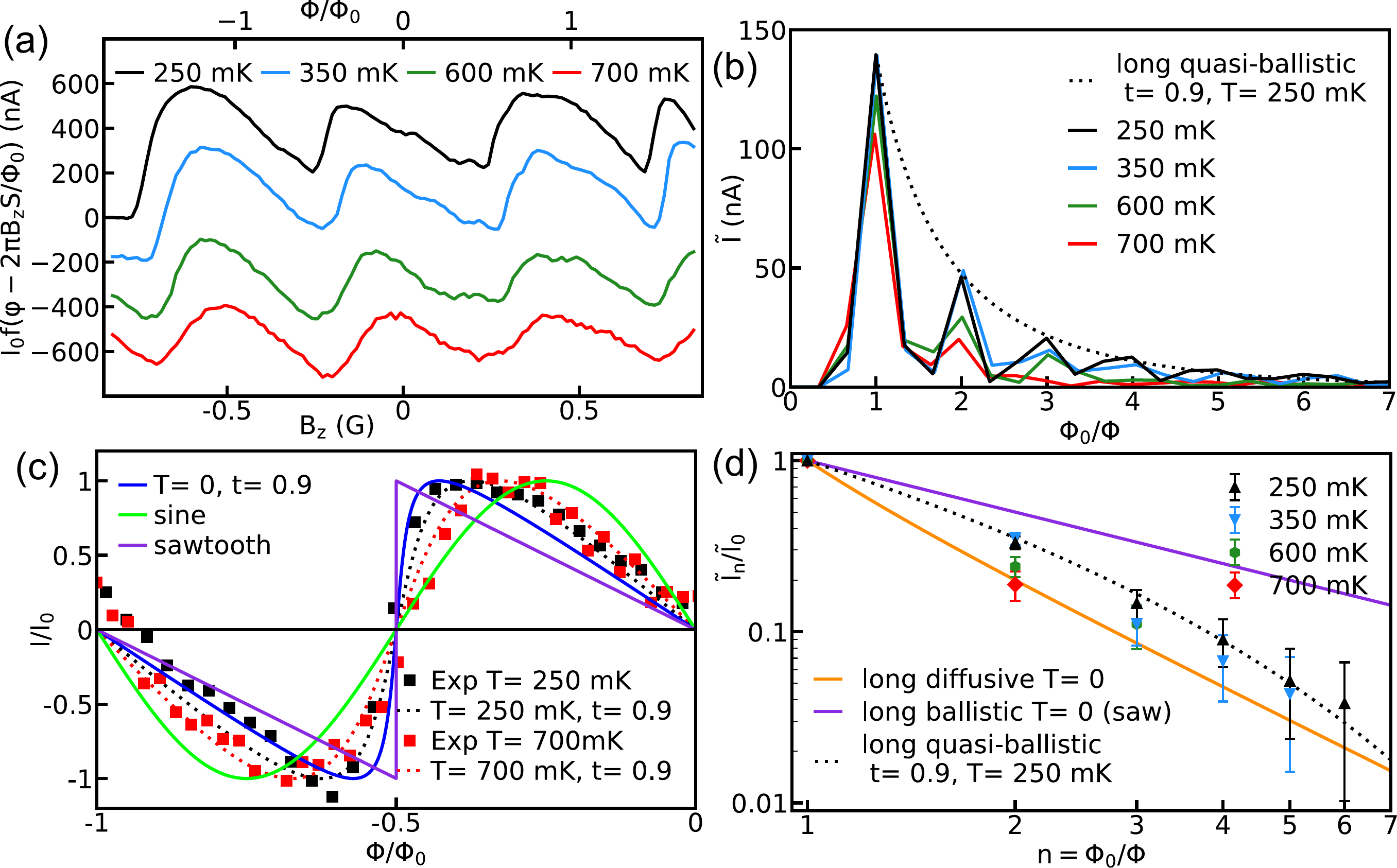}
	\caption{Temperature dependence and harmonics content of SQUID C's  edge junction CPR. (a) CPR at 250, 350, 600 and 700 mK, obtained by high-pass filtering the SQUID's critical current. Curves are shifted vertically for clarity. (b) Fast Fourier Transform (FFT) of curves in (a), and (dashed black line) quasiballistic dependence given by Eq. (2), $\tilde{I}(x)\propto e^{-a(250~mK)x}/{x}, x=\Phi_0/\Phi\geq 1 $, where $a(250~mK)=0.384$. Six harmonics are visible at $250\mathrm{~mK}$, three at $600\mathrm{~mK}$ and two at $700\mathrm{~mK}$.  (c) Symbols: CPR over one period, at 250 and 700 mK, compared to the quasiballistic expressions (1) and (2)  with $t=0.9$ and $E^{edge}_{Th}=6.5~K$(full lines). The quasiballistic result at T=0, rounded solely by the $t=0.9$, is plotted as a purple line. The sawtooth CPR, corresponding to $t=1$ and zero temperature, is plotted in blue. The rounding of the CPR by temperature in the experiment is clear, but the CPR are nonetheless much more skewed than a sinus, shown in green.  (d) Relative amplitude of the FFT peaks, marks on a log-log scale, compared to the zero temperature predictions of the long ballistic CPR FFT (blue line),  Eq. (1) and (2) for $T={250~mK}$, using $t=0.9$ and $E_{Th}^{edge}=560~ \mu eV \simeq 6.5~K$ (black dashed line) \cite{Murani2016,kayyalha2020}, and the long diffusive case, black dotted line. The decay with field of the bulk junction's critical current limits the FFT resolution. We estimate a $\SI{3}{\nano\ampere}$  uncertainty for $\tilde I_{n}$ , see appendix.}
	\label{fig:Fig_2}
\end{figure*}

\begin{figure}[htb]
	\centering
    \includegraphics[width=9 cm]{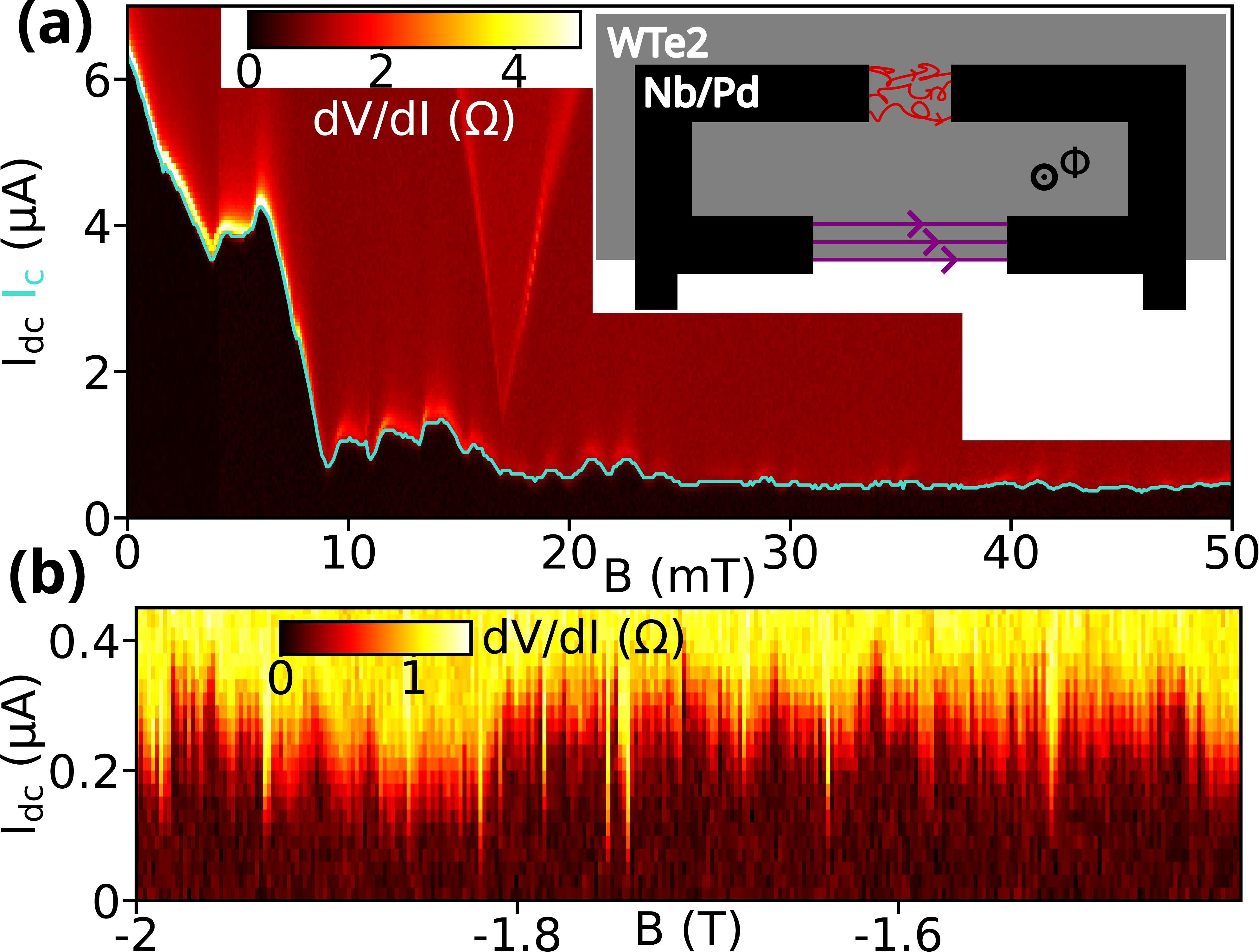}
	\caption{Evolution with magnetic field of SQUID C's interference pattern at T=20 mK. (a) Color-coded differential resistance as a function of dc current $I_{dc}$ and magnetic field $B$ between 0 and 50 mT, and critical current outlined by the turquoise line. The bulk junction is the reference junction in this field range. Its $\mu A$-range critical current decays with a $10~mT$-period Fraunhofer-like behavior, typical of a wide junction. The edge junction's CPR adds a 2 mT-period sawtooth-shaped modulation. $B$ is predominantly along the crystal's $b$ axis. $B_z$, the component along the $c$ axis is $2\%$ of $B$. (b) Interference effects at higher field. The bulk junction's critical current is negligible  and the SQUID's critical current is solely due to the edge junction (a 150 nA current offset has been removed, see appendix).  Inset: Sketch of the asymmetric SQUID C, with the reference junction (diffusive trajectories in red), and the edge junction, here with three ballistic 1D edge states (in purple).   
     }
 	\label{fig:Fig_3}
\end{figure}

Out of the six bulk/edge SQUIDs we have measured, SQUID A displays such a full modulation of $I_c$, Fig.\ref{fig:Fig_1}(a) . SQUIDs B1, B2, D1 and D2 display a small sinusoidal  modulation  of $I_c$, see Fig.  \ref{fig:Fig_1}(b), which disappears in fields of the order of 100 G, see appendix D. This suggests that the junctions with the smaller current have a sinusoidal CPR. SQUID C, with the shortest edge junction, displays a sawtooth-shaped modulation (Fig. \ref{fig:Fig_2}), revealing that the edge junction's CPR is that of a ballistic junction. 
In contrast to $I^{bulk}_c$, which is divided by two on a $B_z$ field scale of 2 G and a temperature scale of 300 mK, $I^{edge}_c$ is practically unchanged up to hundreds of Gauss, and decreases by only $25\%$ between 20 mK and 700 mK (Figs \ref{fig:Fig_2}, \ref{fig:Fig_3} and \ref{fig:Fig_suppmat_IcT})
a robustness that we assign to topological protection. 

\section{ANALYSIS OF THE ASYMMETRIC SQUID WHICH REVEALS BALLISTIC EDGE CONDUCTION}

This SQUID C, shown in Fig. \ref{fig:Fig_1}(e) and (f), is based on a 223 nm-thick WTe$_2$ flake. The superconducting contacts on the edge junction are $600~\mathrm{nm}$ apart, with a lateral overlap on the flake that is smaller than 300 nm, see also Fig. \ref{fig:AFMSQUIDC}. The bulk junction has an hourglass shape, a shape specifically chosen to allow the investigation of high field properties. Figs.  \ref{fig:Fig_1}(c) and \ref{fig:Fig_3}(a) show SQUID C's differential resistance as a function of dc current and field.  A $I_0\simeq 200~\rm{nA}$, $2.35~\mathrm{mT}$-periodic component, with a clear sawtooth shape, is superimposed on a larger (several $\mu\rm{A}$), more slowly varying background, which we identify as $I^{bulk}_{c}$. Despite its unconventional hourglass shape, this bulk junction behaves as a wide and long disordered Josephson junction, see appendix C: $I^{bulk}_{c}(B)$ has the Fraunhofer-like dependence expected of a diffusive junction with an aspect ratio  $W/L=1.5$  \cite{Chiodi2012,Cuevas2007,SM}.  $I^{bulk}_{c}(T)$ also follows the roughly exponential decay expected of a long diffusive junction, $I^{bulk}_{c}(T) \simeq \exp(-T/T_c)$, where the characteristic temperature $T_c$ is related to the Thouless energy via $T_c=3.8 E^{bulk}_{Th}$ \cite{Li2016,Wilhelm}. This yields $T_c\simeq 250 \mathrm{~mK}$, and $E^{bulk}_{Th}\simeq 65~\mathrm{ mK}$, much less than the superconducting gap, which is greater than $2 \mathrm{~K}$, confirming the disordered long junction regime. Finally, combining $I^{bulk}_{c}\approx 7 \mathrm{\mu A}$ with the normal state resistance $R_N\simeq1 \mathrm{~\Omega}$ yields $eR_NI^{bulk}_{c}= 80~\rm{mK} $, four to ten times less than the perfect interface value $eR_NI_c^{bulk}=10.82E^{bulk}_{\rm{Th}}$, indicating that the transparency of the NS interface  is less than $1$. More details on the bulk junction are given in Appendix C.

\begin{figure}[htb]
	\centering
	\includegraphics[width=9cm]{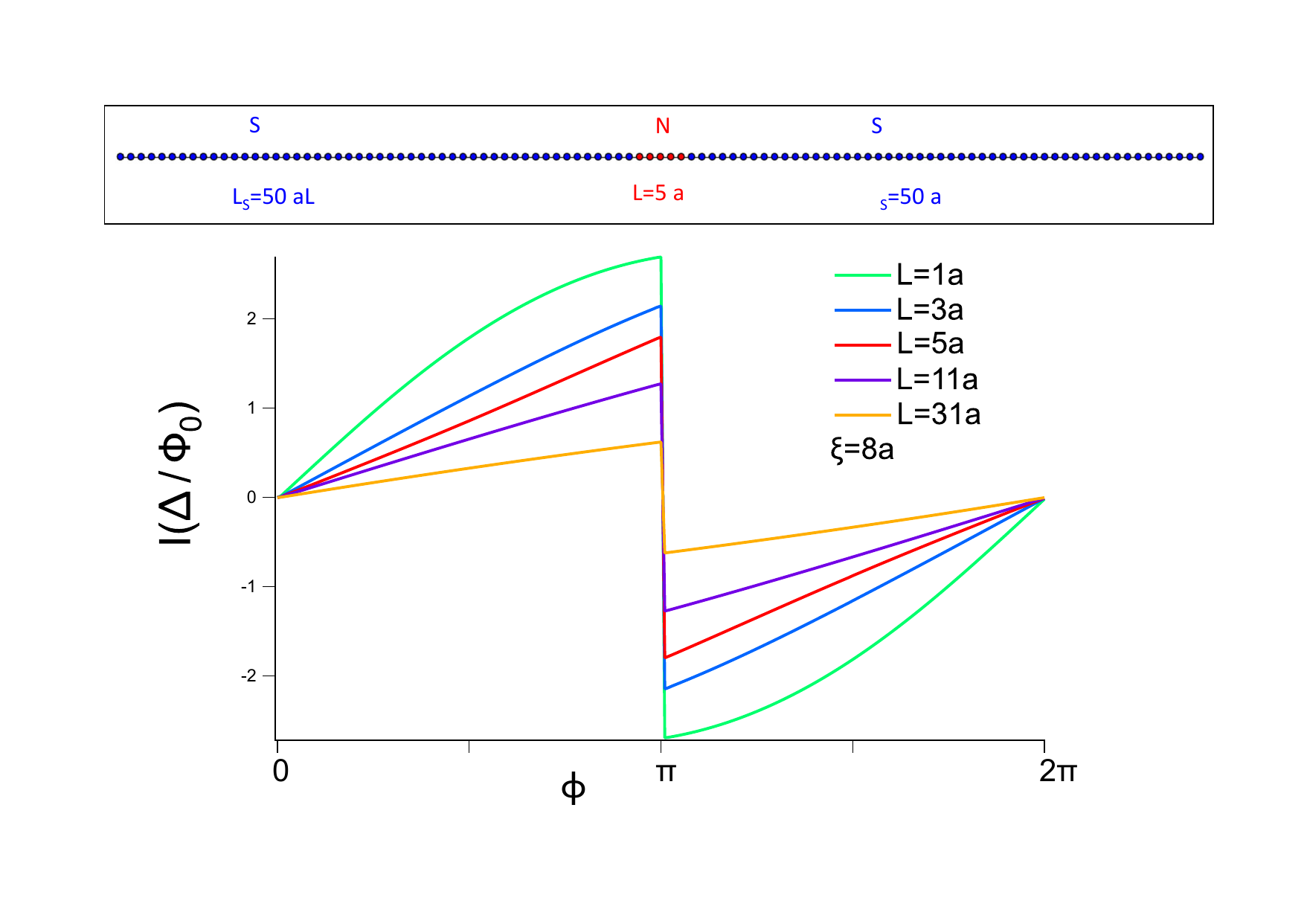}
	\caption{Tight-binding simulation of a  ballistic 1D SNS chain with varying number of normal sites, and its zero-temperature Supercurrent-versus Phase relation (CPR). One can notice that the linear, sawtooth CPR of a so-called long ballistic junction is achieved for a normal chain length longer than 3 sites, well below the long junction limit of $\xi\gg L$. The CPR made of branches of sinus, the so-called short ballistic junction CPR is only verified for a one-atom long chain. In this calculation the hopping energy is 4, superconducting gap $\Delta=1$, Fermi energy $E_F=0$, the superconducting reservoirs are each 50 sites long. The superconducting coherence length is 8 sites long, $\xi=8a$, and the disorder on on-site energy is taken to be $10^{-4}$. L is the length of the normal chain,  L=1a, 3a, 5a, 11a and 31a, with a the interatomic distance in the chain.}
	\label{fig:IJshort_long}
\end{figure}

\begin{figure}[bt]
	\centering
	\includegraphics[width=8cm]{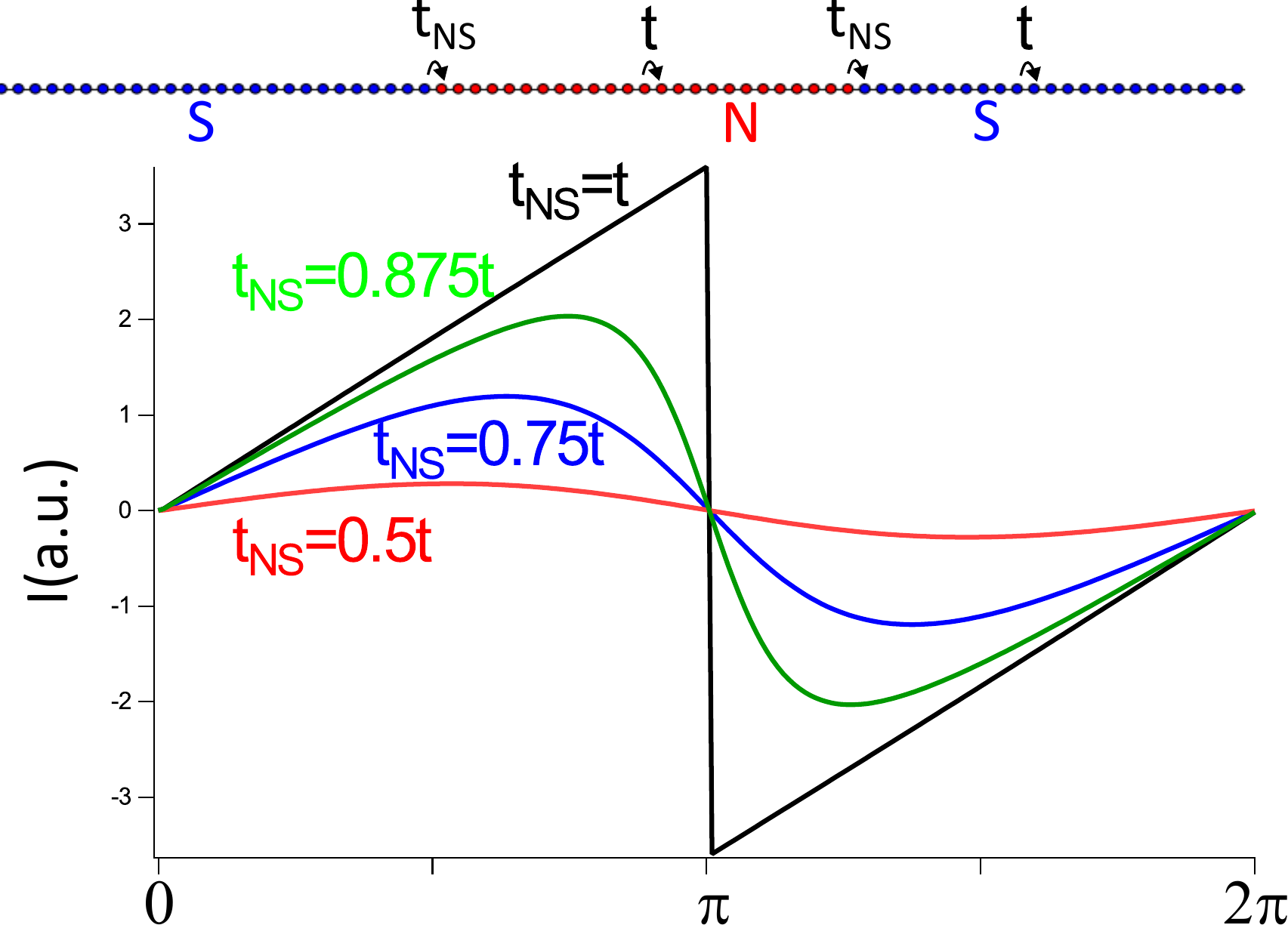}
	\caption{Non topological 1D Josephson junction without disorder, for varying interface transmission $t_{NS}$. As sketched, the tight binding model contains two 1D, disorderless superconducting electrodes 50 sites-long, one 1D disorderless N chain with 25 sites. The Fermi energy=0, nearest neighbor hopping $t=4$, $\Delta=1$, and $t_{NS}/t$ varies between 1 and 0.5. The jump at phase difference $\pi$ is rounded as soon as the interface transmission is not perfect, in contrast with the topological junction case, see figure below. 
    }
	\label{fig:Fig_suppmat_tNS_nontopo}
\end{figure}

The edge junction has strikingly different characteristics. Its supercurrent is easier to extract above $\SI{250}{\milli\kelvin}$, when $I^{bulk}_{c}$ is small enough that its field dependence is not a hindrance (see appendix).
We find that $I^{edge}_{c}$ is much less sensitive to temperature and magnetic field than $I^{bulk}_{c}$:
$I^{edge}_c$ decreases by only 25\% between 20 mK and 700 mK, see Fig. \ref{fig:Fig_2}, and is almost unaffected by magnetic fields up to hundreds of Gauss, $I^{edge}_{c}\simeq I_0\simeq 200 \mathrm{~nA}$, see appendix. 
Note that above those temperatures and fields, $I^{bulk}_{c}$ is too small to consider that the SQUID's $I_c$ reflects the CPR of the edge junction, as discussed later. This very slow decrease of $I^{edge}_{c}(T)$  is a first indication of a much greater Thouless energy and thus a cleaner regime than the bulk junction. We compare $I^{bulk}_{c}(T)$ and $I^{edge}_{c}(T)$ in Fig. \ref{fig:Fig_suppmat_IcT}.

\begin{figure}[tb]
	\centering
	\includegraphics[width=9cm]{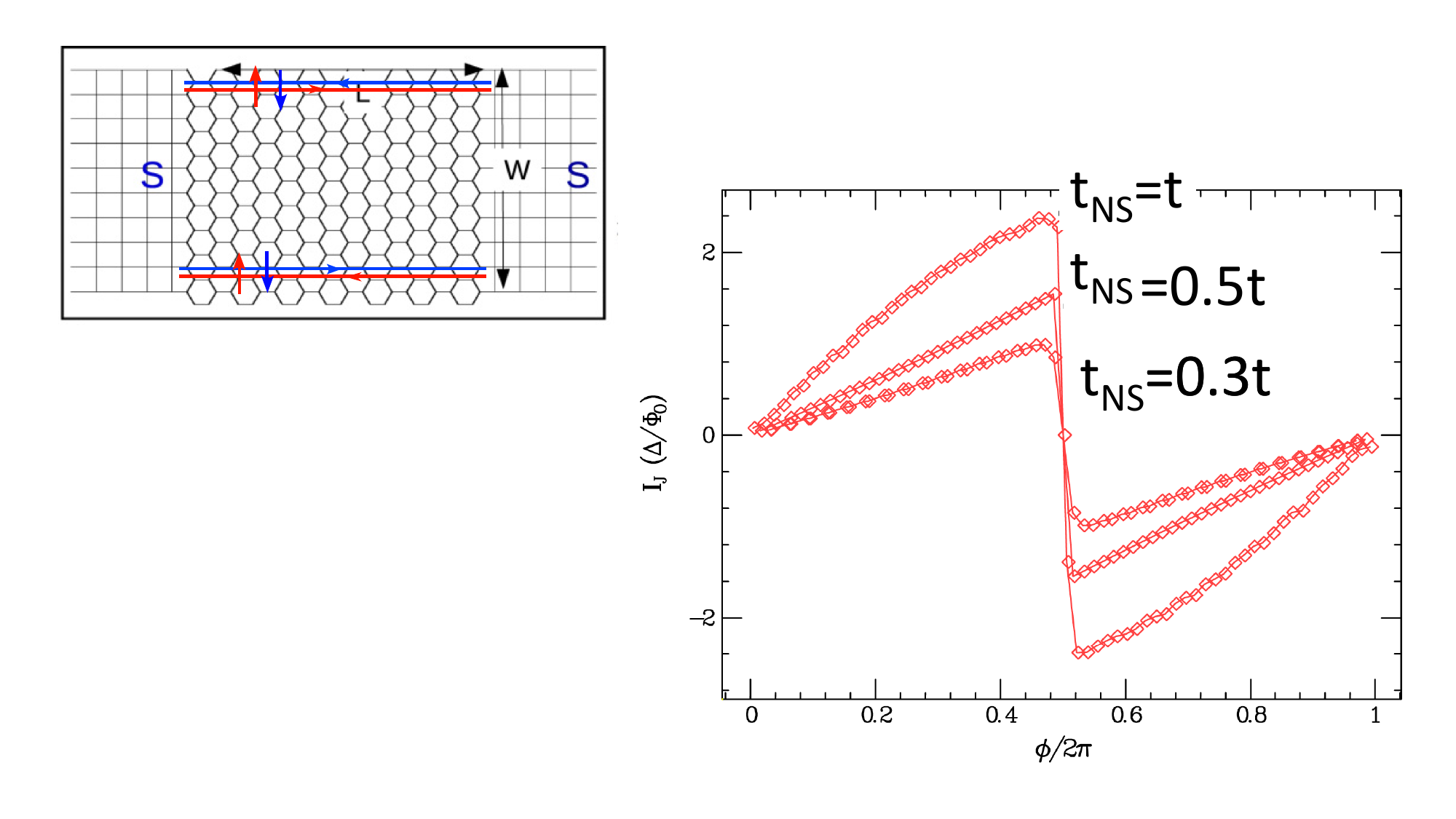}
	\caption{Topological 2D Josephson junction with small disorder, realized with the Kane-Mele model, as presented in detail in \cite{murani_andreev_2017}.  The CPR of the Quantum Spin Hall 1D helical states is presented for varying interface transmission $t_{NS}$. We find that, in contrast to the 1D chain presented above, that has no disorder but no topological protection, in the Quantum Spin Hall case the  jump at phase difference $\pi$ is not  rounded when the NS interface transmission decreases. Only the amplitude of the jump decreases. This is similar to what was published in \cite{murani_andreev_2017,cayao_andreev_2018}.
    }
	\label{fig:Fig_suppmat_tNS_topo}
\end{figure}

\begin{figure}[tb]
	\centering
	\includegraphics[width=9cm]{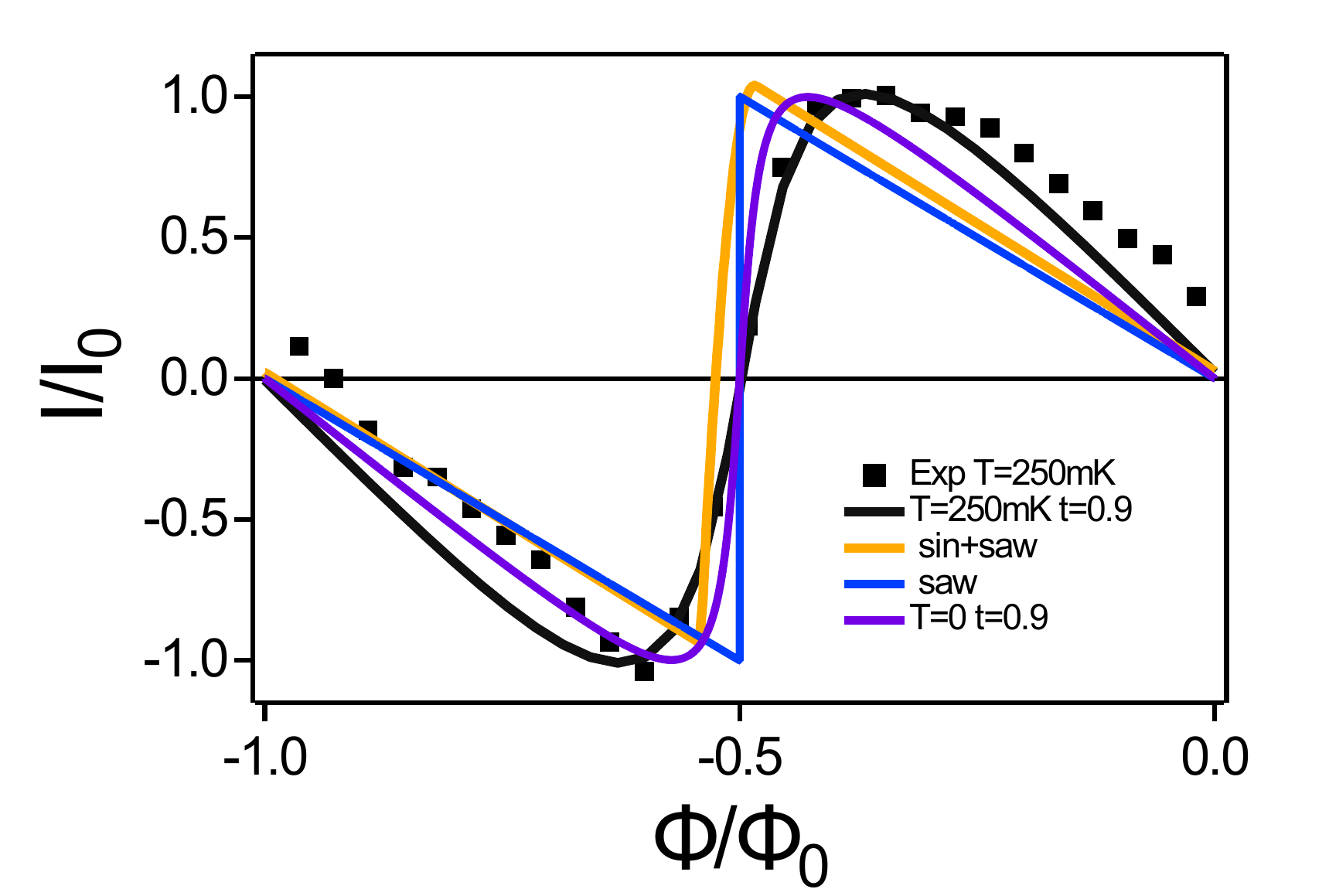}
	\caption{Different causes for the CPR appearing rounded. This figure compares: i) the sharp sawtooth result, in blue, which is the CPR of a long ballistic junction at T=0; ii) The quasiballistic CPR in which the rounding is given by the coefficient $t=0.9$, see main text. The CPRs are shown at T=0 in purple and T=250 mK in black, for an $E_{Th}=6.5 ~K$, along with the T=250 mK experimental CPR shown as black squares; and iii) in orange the critical current expected of an asymmetric SQUID in which the reference junction has a sinusoidal CPR and the weak link has a sawtooth CPR with an amplitude 30 times smaller than the reference junction, as in the experiment. The critical current reflects the sawtooth CPR of the weak link over most of the flux region, but in a (small) flux region around which the sawtooth jumps,  the critical current varies like the CPR of the reference junction rather than the CPR of the weak link junction. 
    }
	\label{fig:Fig_suppmat_sinsaw}
\end{figure}

We now focus on the edge junction's CPR. It is extracted by high-pass filtering the SQUID's critical current $I_c(B)$, and plotted in Fig. \ref{fig:Fig_2} (a) for four temperatures. A sawtooth shape is clearly visible, far different from the sinusoidal CPR of conventional Josephson junctions. 
Up to six harmonics are visible in the CPR's FFT at low temperature, see Fig. \ref{fig:Fig_2}(b). Each harmonic is plotted at different temperatures in  Fig. \ref{fig:Fig_2}(d), and compared to the predicted zero temperature ballistic and diffusive limits. We find that the edge junction's CPR harmonics decay more slowly than the CPR of a diffusive junction, confirming that the edge states are not in a diffusive regime. We argue that the high harmonics content and its robustness is an indication of topological protection, by contrasting the rounding (by an imperfect interface transmission) of the CPR of topologically protected conductors and nonprotected conductors. Let us first consider a Josephson junction with a perfect conductor (no backscattering) as the weak link. The junction's CPR features a jump at phase difference $\varphi=\pi$ (see Fig. \ref{fig:Fig_2}(c) and Fig. \ref{fig:IJshort_long}). 
This jump translates into a high harmonics content, with the amplitude of $n$ -th harmonic decaying as $1/n$, see Fig. \ref{fig:Fig_2}(d). In the non-topologically protected case, backscattering within the weak link as well as at the NS interface cause rounding of the CPR and decrease of the critical current. The harmonics decay faster (as $1/n^2$ in a diffusive junction, see \ref{fig:Fig_2}(d) and  Fig. \ref{fig:Fig_suppmat_tNS_nontopo}).
By contrast, for a topological junction, it has been shown \cite{adroguer_probing_2010,cayao_andreev_2018,kwon_fractional_2004,
murani_andreev_2017,SM} that an imperfect interface transmission
does not round the CPR, but merely decreases the supercurrent amplitude (Fig. \ref{fig:Fig_suppmat_tNS_topo}).
Therefore, a jump in the CPR at $\pi$ is strongly suggestive of topological states. Of course, the CPRs of topological materials can be rounded by several factors such as interactions, coupling between helical states with opposite helicities, or the junction's electromagnetic environment. We also note that a CPR with a discontinuity will also be rounded in an asymmetric SQUID measurement, as pointed out by \cite{babich2023,endres2023,alexandrebernard2022}, and illustrated in Fig. \ref{fig:Fig_suppmat_sinsaw}. 

Quantitatively, the rounding of the CPR by temperature $T$, which translates into a stronger $T$-decay of the higher harmonics, can be analysed to estimate the edge junction's Thouless energy $E^{edge}_{Th}$ as well as the zero temperature, intrinsic (i.e. without contacts) transmission $t$ through the edge state. $t$ characterises a small avoided crossing of the Andreev levels at $\pi$ near zero energy,  causing a zero temperature rounding of the CPR. Both temperature and $t$ contribute to the decay of the $n_{th}$ harmonic $\tilde I_{n}$, according to: 
\begin{equation}
I^{edge}(\varphi)=\sum_n \tilde I_n \sin{n \varphi}=I_0\sum_n e^{-a(T)n} \frac{(-1)^n}{n} \sin{n \varphi},
\end{equation} with the damping coefficient $a(T)$ relates to $t$ via \cite{Murani2016}
\begin{equation} 
\begin{split}
e^{-a(T)n} & = t^ne^{-n2\pi k_BT/E^{edge}_{Th}} , i.e.\\
a & = 2\pi k_BT/E^{edge}_{Th}-\ln(t).
\end{split}
\end{equation}
Fitting at each $T$ the ratio $\tilde I_{n}/\tilde I_{1}=e^{-a(n-1)}/n$   for $n \in [1, 6]$ with $a$ the only fitting parameter yields $E^{edge}_{Th}=560\pm75\mathrm{~\mu eV}$ and $t=0.89\pm0.04$, see Appendix C. Thus $E^{edge}_{Th}$ is two orders of magnitude greater than $E^{bulk}_{Th}$, i.e. an order of magnitude more than expected if both junctions were in the same (long diffusive) regime (given the factor three difference in the junction lengths). The high $E^{edge}_{Th}$ is thus a clear indication that the edge junction is in a quasi-ballistic regime. In fact, $E^{edge}_{Th}$ determined this way agrees with the ballistic expression $E^{edge}_{Th,ball}=\hbar v_F/L^{edge}\simeq520~\mu eV$, given the edge junction length $L^{edge}=600\mathrm{ ~nm}$ and the estimated Fermi velocity of 1D states $v_F\simeq 5\times 10^5 \mathrm{~m.s^{-1}}$ \cite{zhao2021}. The high value of $t$ is also suggestive of a topological protection. The fact that $t\simeq0.9$ rather than $1$ may be attributed to a small coupling between helical states a distance $W$ apart, which causes an avoided crossing of Andreev levels at $\pi$ that is proportional to $\exp{-W/\xi}$, with $\xi$ the superconducting coherence length \cite{murani_andreev_2017}. Evaluating $\xi$ in the ballistic junction regime yields $  \xi=\frac{\hbar v_F}{\pi \Delta}\approx 200-600 \mathrm{~nm}$, for $\Delta \simeq 170-500~ \mu eV$ (the precise value of $\Delta$ at the contact is hard to determine, see Appendix). Thus a coupling between edge states can be significant if they are less than 100 nm apart, as is the case for some of the step edges, see Fig. \ref{fig:Fig_1}(f) and \ref{fig:AFMSQUIDC}. The coupling of helical states to trivial (nonhelical) states would also round the CPR and thus decrease $t$.

We now discuss the number of helical states that carry the edge supercurrent. The expected critical current of a single helical state \cite{Beenakker2013} in the respectively long or short junction regime is:
$$i^{long}_{ch}=\frac{ev_F}{2L^{edge}}=\frac{eE^{edge}_{Th}}{2\hbar}\approx 68 \pm9 \mathrm{~nA}$$
$$i^{short}_{ch}=\frac{e\pi\Delta}{h}\approx 20-60 \pm 10\mathrm{~nA}$$
for $\Delta \simeq 170-500~ \mu eV$.
Thus the total supercurrent through the edge junction, $200$ nA, must be carried by three of more helical states, depending on the NS interface quality. Such states could be located at the step edges visible in the sample images of Fig. \ref{fig:Fig_1} and \ref{fig:AFMSQUIDC}. We note that the edge junction length $L^{edge}= 600~\mathrm{~nm}$ is between 1 and 3 times $\xi$, corresponding to an intermediate regime, between short and long, that is characterized by a sawtooth shaped CPR (see appendix). The exact number of channels as well as the interface transmission distribution is difficult to determine. More helical channels are required to carry the supercurrent if the NS interface transmission ${t_{NS}}$ is smaller than $1$, since the supercurrent carried by each helical channel is reduced (roughly as the transmission amplitude $\sqrt{t_{NS}}$, in contrast to the stronger ${t_{NS}}$ reduction of non helical channels \cite{kwon_fractional_2004,BrinkmanGolubov2000}). We can however posit that the number of channels is not too large, given the interference pattern we measure: a large number of channels would lead to scrambled interference at relatively low magnetic field, in contrast to the measurement, as we now discuss. 

With increasing field the bulk and edge junctions swap roles (Fig. \ref{fig:Fig_3}). At low field the bulk junction acts as the reference junction,
due to its order-of-magnitude larger critical current, and the modulation of the SQUID's $I_c$ is related to the edge junction's CPR. However, a relatively low field of a few tens of milliTesla causes $I_c^{bulk}$ to decay in an oscillatory, Fraunhofer-like manner, because of interference between diffusive trajectories within the bulk junction \cite{Cuevas2007}. By contrast, $I_c^{edge}$ is unaffected by field and becomes the largest current above $100~\rm{mT}$. The SQUID's $I_c$ is then in fact dominated by $I_c^{edge}$, and the CPR of the bulk junction is expected to modulate $I_c$, a small modulation that is not visible (see appendix). One can notice an aperiodic modulation on the scale of a few hundred mT at high field, see Fig. \ref{fig:Fig_3} (b). Such modulations must be due to interference between edge channels. Since two channels would cause more regular interference, we believe that at least three channels are likely (see Appendix C for a discussion of interference patterns in different edge states configurations). Such interference could occur due to orbital dephasing by the in-plane field, generating a flux between states on the edges of different atomic steps. It could also be due to Zeeman dephasing by different g factors, given the large in-plane field.

Finally, the lateral extension of the edge states can be estimated by noting that the critical current decay due to orbital dephasing by the magnetic field occurs on a scale of a flux quantum through the edge state area. The $3\ $T-scale in plane, and the $0.1\ $T-scale along the $c-$axis suggest that the hinge states extend over $2\ $nm in the $ac$-plane and $30\ $nm in the $ab$-plane, demonstrating the 1D character of the hinge states. 

\section{CONCLUSION}
To conclude, we have probed six junctions at the edges of high quality multilayer WTe$_2$ flakes. The shortest, 600 nm-long junction displays a strongly non-sinusoidal CPR, with a steep slope at $\pi$ that is characteristic of quasi-ballistic channels. At least three channels carry the supercurrent, in magnetic fields up to 3 Tesla, pointing to narrow edge states that could be situated at the atomic step edges of the flake. These results support the hypothesis of a SOTI character of WTe$_2$, as expected given its double band inversion and screw-symmetry. The sinusoidal CPR found for some SQUIDs can be explained by several factors: edge states may simply not systematically be contacted by the superconducting contacts, so that the sinusoidal CPR would be that of disordered trivial bulk states. Even if contacted, the edge states in those other, longer edge junctions would carry smaller critical currents, which could cause a sizable rounding of their CPR by a dissipative environment such as the trivial bulk states of WTe$_2$. The fact that helical edge states can be found in multilayer samples, which are easier to fabricate than WTe$_2$ monolayers, make multilayer WTe$_2$ interesting candidates for the exploration of 1D helical states, for instance via microwave spectroscopy.

\section{ACKNOWLEDGMENTS}
 We acknowledge technical help from S. Autier-Laurent and R. Weil, advice on sample fabrication by Sanfeng Wu, discussions with Benjamin Wieder about the topological aspects of WTe$_2$, Arthur Veyrat for help with the manuscript. This work was funded by the European Union through the BALLISTOP ERC 66566 advanced grant. Work at Princeton was supported by NSF through the Princeton Center for Complex Materials, a Materials Research Science and Engineering Center DMR-2011750, the Gordon and Betty Moore Foundation (EPiQS Synthesis Award) through grant GBMF9064, and the David and Lucile Packard Foundation.

\section{DATA AVAILABILITY}
The data that support the findings of this article are openly available \cite{Zenodo}.

\renewcommand{\thesection}{\hspace*{-1 em}}
\setcounter{secnumdepth}{4} 
\renewcommand{\thesubsection}{\arabic{subsection}}


\section{Appendix A: Considerations on the Shape of the CPR in different situations}
\subsection{Transition from short to sawtooth-shaped CPR}
As displayed in Fig. \ref{fig:IJshort_long}, the sawtooth-shaped CPR characteristic of a long ballistic junction in fact develops in rather short junctions, even shorter than the superconducting coherence length, and is not restricted to the regimes $\xi\gg L$. In fact, it is the so-called short-junction CPR that is rarest, and restricted to junctions just a few atoms long.

\begin{figure}[tb]
	\centering
	\includegraphics[width=\columnwidth]{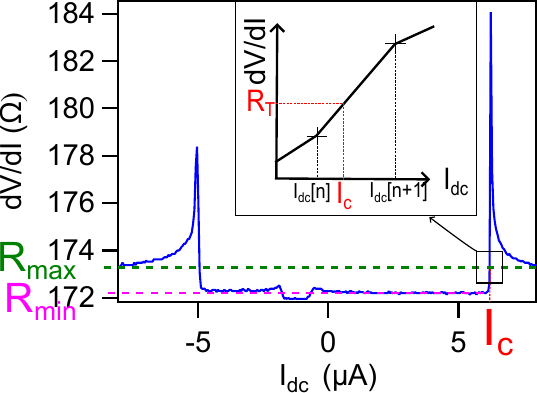}
	\caption{Differential resistance dVdI as a function of dc current for an upward and a downward sweep at $B=0$ and at the lowest temperature, around 20 mK.}
	\label{fig:Fig_suppmat1}
\end{figure}

\begin{figure}[tb]
	\centering
	\includegraphics[width=\columnwidth]{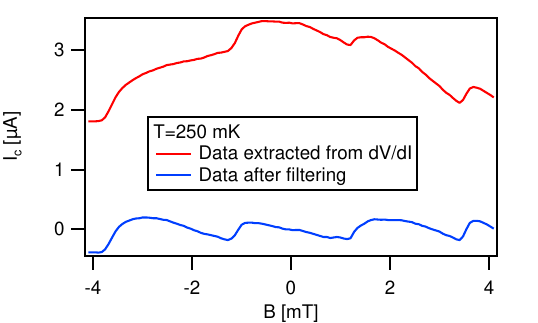}
	\caption{Red: Variations with field of the "raw" critical current extracted from the differential resistance as a function of dc current, according to the procedure described in the text. Blue : CPR obtained after filtering of the raw critical current curve.}
	\label{fig:Fig_suppmat2}
\end{figure}

\begin{figure}[tb]
	\centering
	\includegraphics[width=\columnwidth]{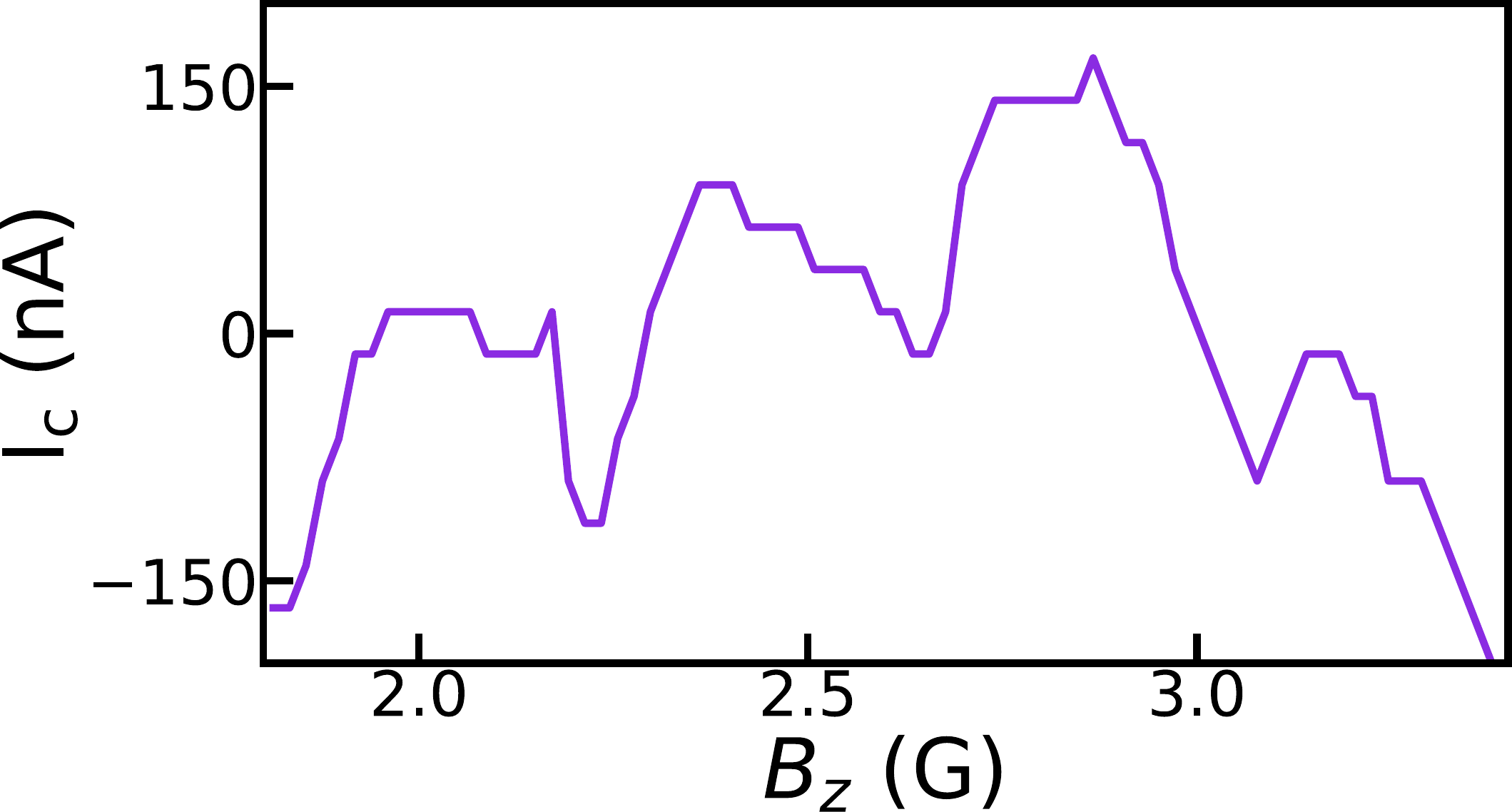}
	\caption{Extracted Current phase relation at 20 mK, the lowest temperature. The extraction of this curve is complicated by instabilities due to superconducting vortices around zero field, so the CPR was extracted from the SQUID's critical current at higher field for which the critical current of the reference junction is much smaller. The quality of these data is much worse than, yet consistent with the higher temperature data. }
	\label{fig:Fig_suppmat_CPR20mK}
\end{figure}

\begin{figure}[tb]
	\centering
	\includegraphics[width=\columnwidth]{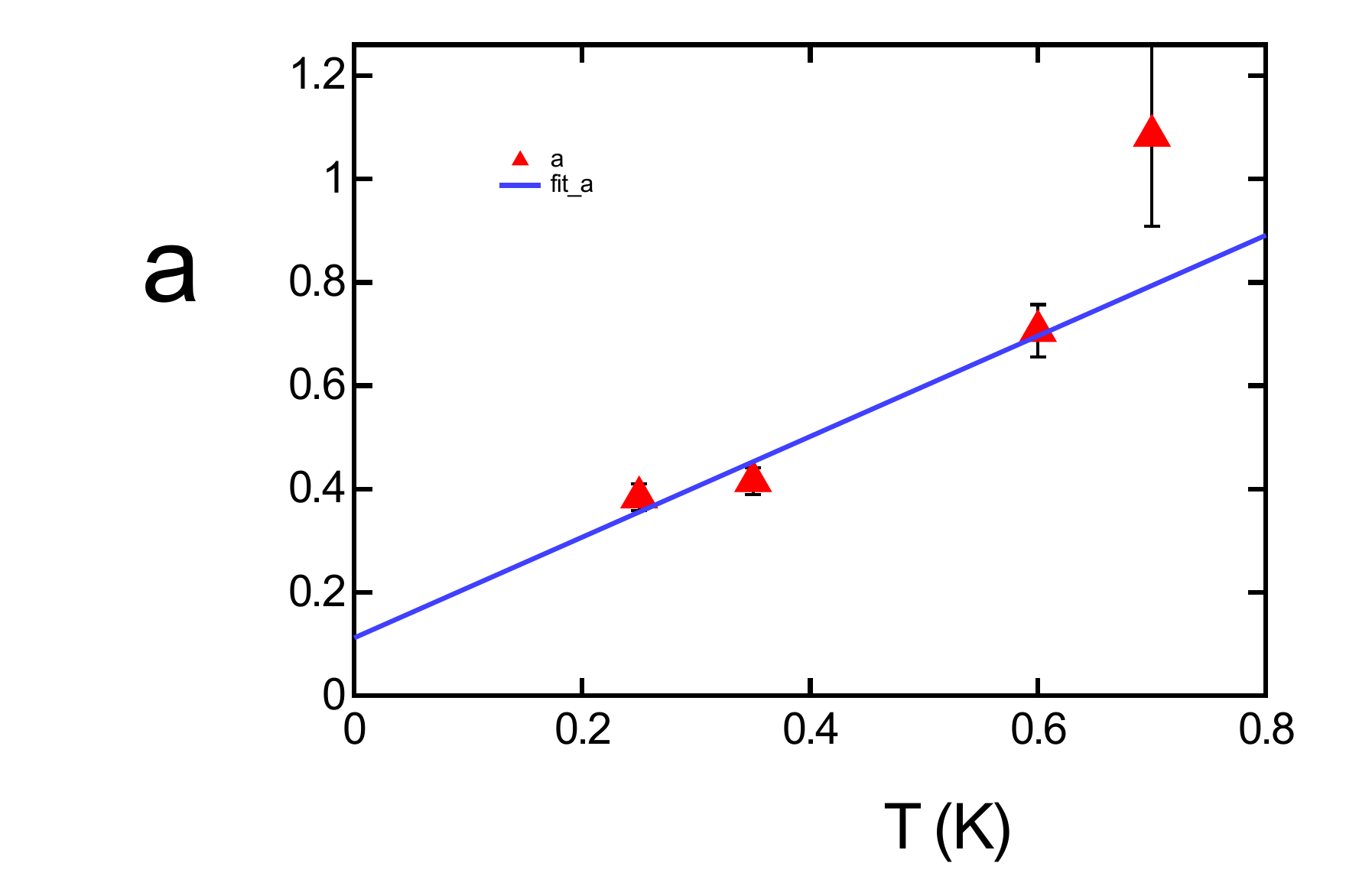}
	\caption{ Extracting SQUID C's edge junction Thouless energy and intrinsic transmission factor $t$, from the decay of the CPR's harmonics's relative amplitudes  expressed as a function of the decay coefficient $a$ via $\frac{\tilde I_{n}}{\tilde I_{1}}=\frac{e^{-a(n-1)}}{n}$. $a$ depends on temperature, and the linear fit to $a=2\pi k_BT/E^{edge}_{Th}-\ln(t)$ yields the edge's Thouless energy $E^{edge}_{Th}=560\pm75\mathrm{~\mu eV}\simeq ~6.5~K$ and the intrinsic transmission coefficient accounting for the CPR rounding $t=0.89\pm0.04$.}
	\label{fig:Fig_suppmata}
\end{figure}

\begin{figure}[tb]
	\centering
	\includegraphics[width=\columnwidth]{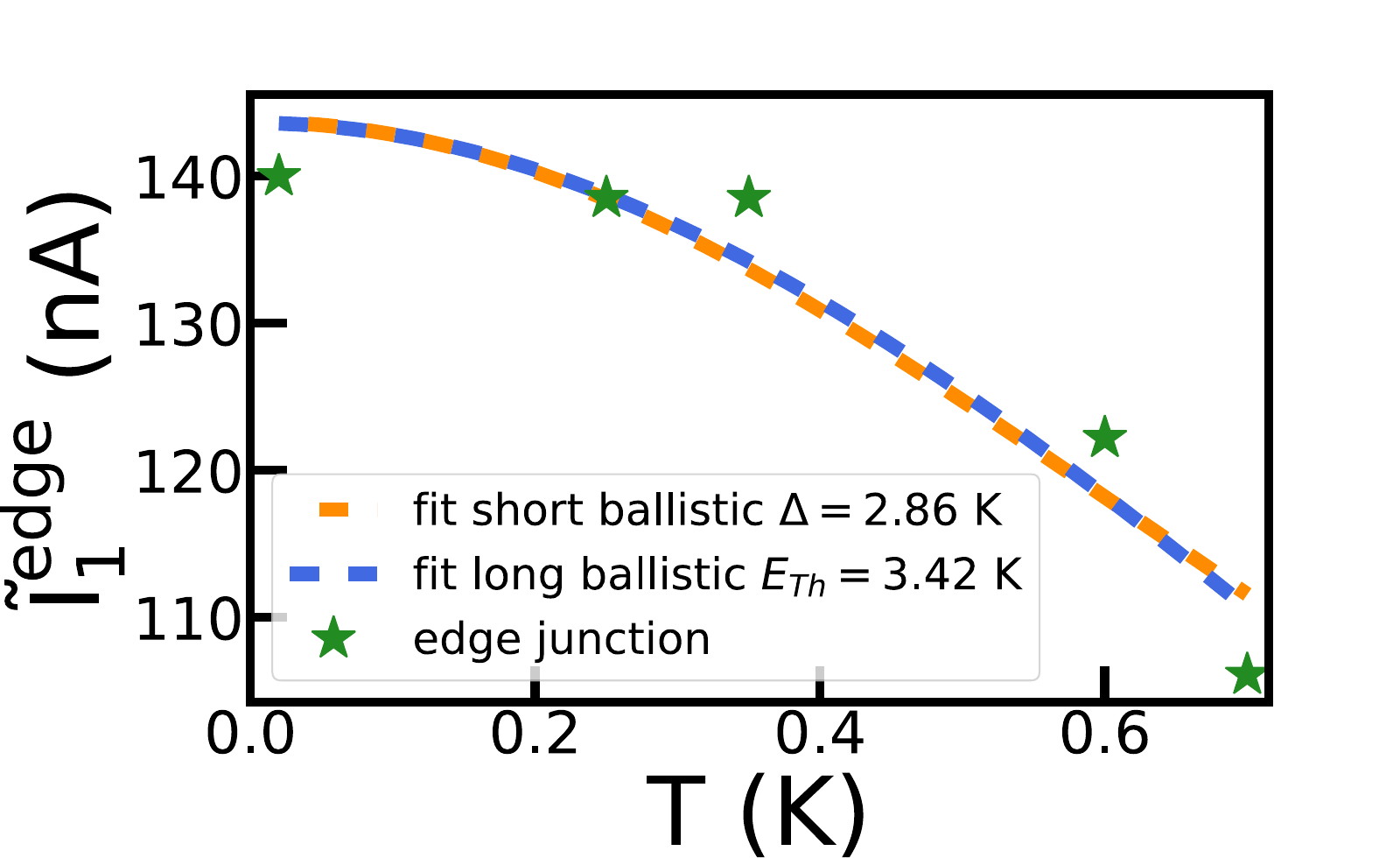}
	\caption{Analysis of the temperature dependence of the critical current of SQUID C's edge junction. Data points are the amplitude of the CPR's first harmonic at different temperatures $\tilde I^{edge}_1$.  We find that the data can be fitted equally well by the temperature dependence of the first harmonic of the ballistic short junction, $I_{short}^{ball}(\varphi,T) \propto \Delta\sin(\frac{\varphi}{2})\tanh(\frac{\Delta}{2T}\cos(\frac{\varphi}{2}))$, or the first harmonic of the ballistic long junction $I_{long}^{ball}(\varphi,T) \propto \sum_{k=1}^{k=\infty}\sin(k\varphi)\frac{4(-1)^{k+1}T}{\sinh(2\pi k\frac{T}{E_{\mathrm{Th}}})}$ \cite{Houzet,Cayssol}.
    The gap energy extracted of the fit to the short junction expression is $\mathrm{\Delta\simeq 2.9~K}$, whereas the Thouless energy extracted from the fit to the long junction expression is $\mathrm{E_{Th}\simeq 3.4~K}$. These results are compatible with the Thouless energy extracted from the analysis of the sawtooth harmonics discussed in the main text, and confirm that the edge junction is in a ballistic regime that is intermediate between short and long.  }
	\label{fig:Fig_suppmat_shortorlong}
\end{figure}

\begin{figure}[tb]
    \centering
    \includegraphics[width = \columnwidth]{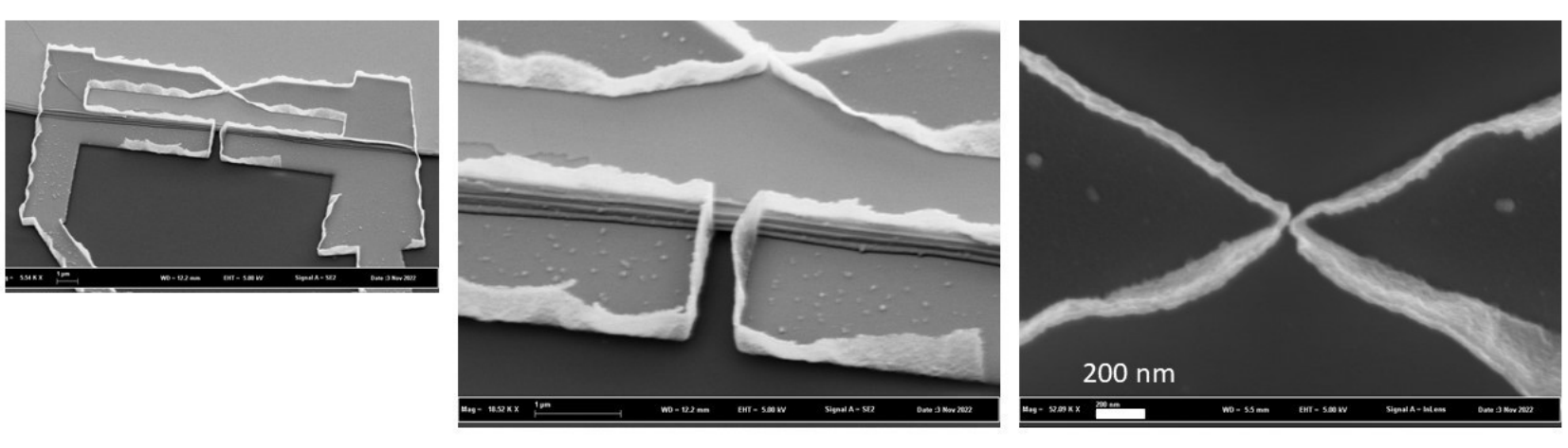}
    \caption{Scanning Electron Micrograph of SQUID C, including zoomed-in views of the bowtie-shaped bulk junction. The white contours of the electrodes are a (common) consequence of the sputter deposition through a resist mask. 
    }
    \label{fig:SEMbowtie}
\end{figure}

\begin{figure}[tb]
    \centering
    \includegraphics[width = \columnwidth]{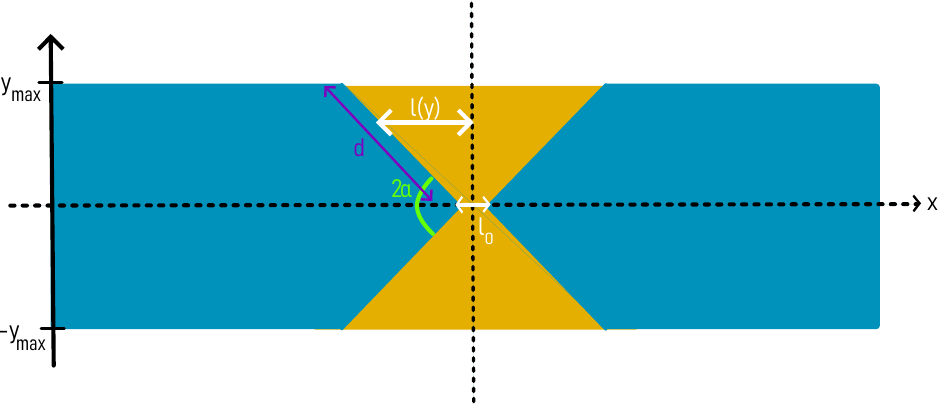}
    \caption{Sketch of the hourglass-shaped junction, with parameters used to determine the junction's conductance. The blue part represents the superconducting leads and the yellow part the WTe2 flake in the junction. $y_{max} = 0.73$ µm, $d = 2.21$ µm. $t = 223$ nm the thickness of the flake, and $l(y_{max}) = 2.06$ µm.}
    \label{fig:triangle}
\end{figure}

\begin{figure}[tb]
    \centering
    \includegraphics[width = \columnwidth]{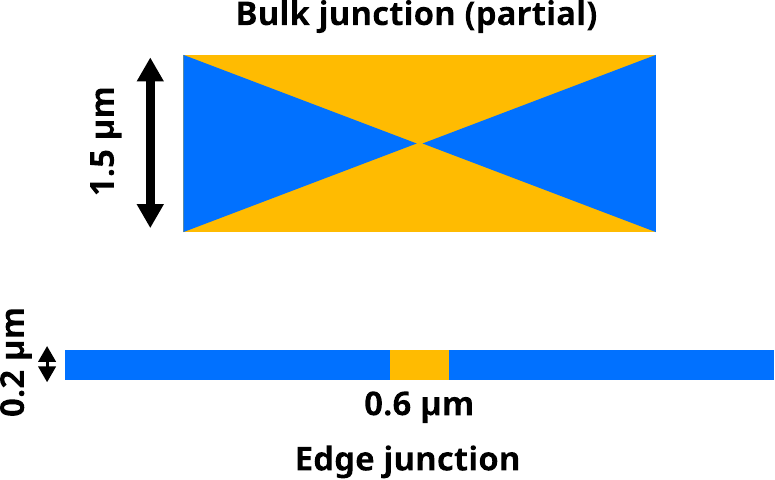}
    \caption{Comparison between the bulk and edge junction geometries, which illustrates why the bulk junction carries the largest supercurrent and therefore must be the reference junction (superconducting Pd/Nb in blue, WTe2 in yellow): The edge junction’s superconducting electrodes are 600 nm apart, with a narrow region of WTe2 as the weak link (bottom sketch). The bulk junction (top sketch) has electrodes with a separation much smaller than 600 nm, and may also have a central PdTex superconducting filament in parallel: the critical current of this bulk junction must be much larger than the edge junction’s. Consequently the bulk junction must act as the reference junction, and the small, 150 nA oscillation must be due to the edge junction.
    }
    \label{fig:comparison_edge_bulk_junctions}
\end{figure}

\begin{figure}[tb]
    \centering
    \includegraphics[width = \columnwidth]{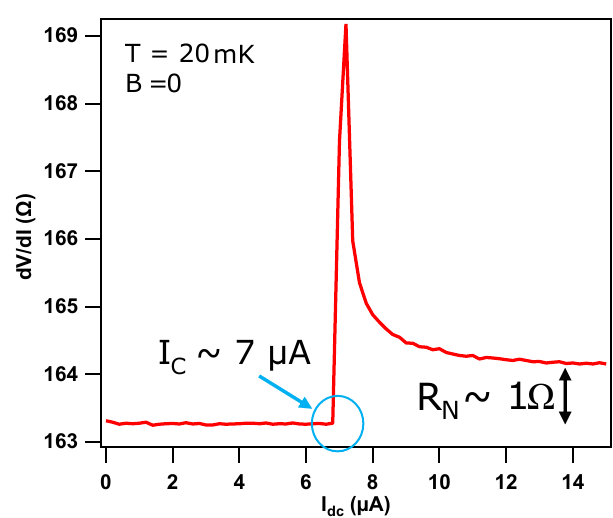}
    \caption{Differential resistance curve at 0 field and T = 20 mK. The critical current, highlighted by the blue circle, is about 7 \textmu A. The jump in resistance, i.e. the difference between the low bias and high-bias region is the sample resistance, about 1 $\mathrm{\Omega}$. The 161.3 $\mathrm{\Omega}$ resistance corresponds to the measurement lines leading to the sample from room temperature.}
    \label{fig:dVdI_single}
\end{figure}

\begin{figure}[tb]
    \centering
    \includegraphics[width = \columnwidth]{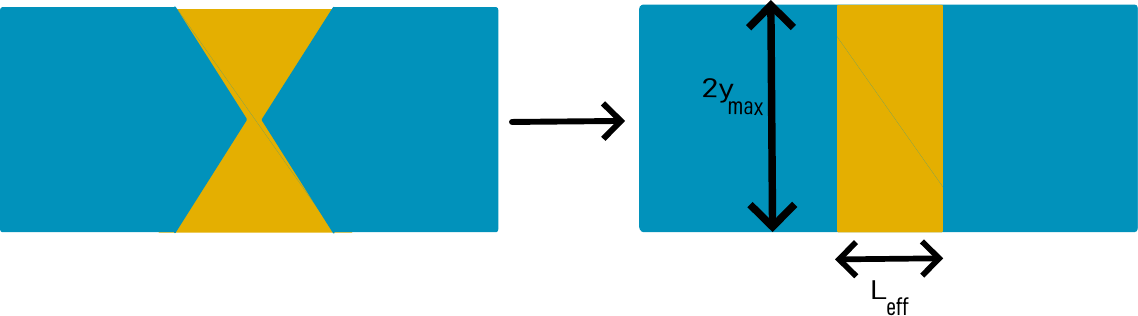}
    \caption{The hourglass junction can be recast as an equivalent rectangular junction of width $w = 2y_{max} \simeq 1.46 ~\mu m$ and length $L_{\rm{eff}} \simeq 0.94~ \mu m$, much simpler to study.}
    \label{fig:transformation_hourglass}
\end{figure}

\begin{figure}[tb]
	\centering
	\includegraphics[width=\columnwidth]{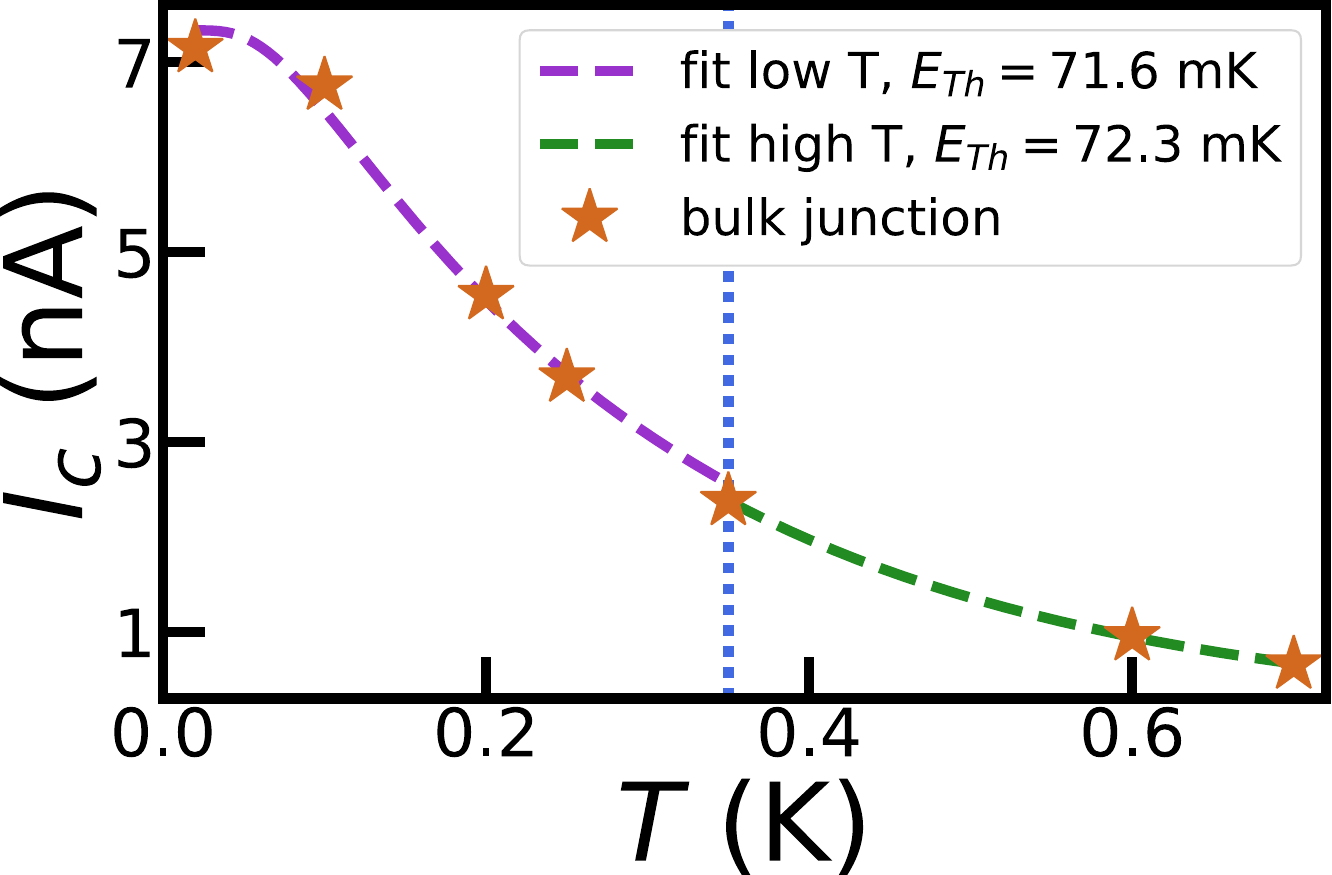}
	\caption{The critical current of SQUID C's bulk junction can be fitted in two parts, using the expression at low temperature ($T<5E_{\rm{Th}}$) and the one at high temperature ($T>5E_{\rm{Th}}$), yielding the same $E_{\rm{Th}}\approx$ 70 mK and a cutoff at $5E_{\rm{Th}}\approx$ 350 mK.}
	\label{fig:Fig_suppmat_longdiffusive}
\end{figure}

\begin{figure}[tb]
	\centering
	\includegraphics[width=8cm]{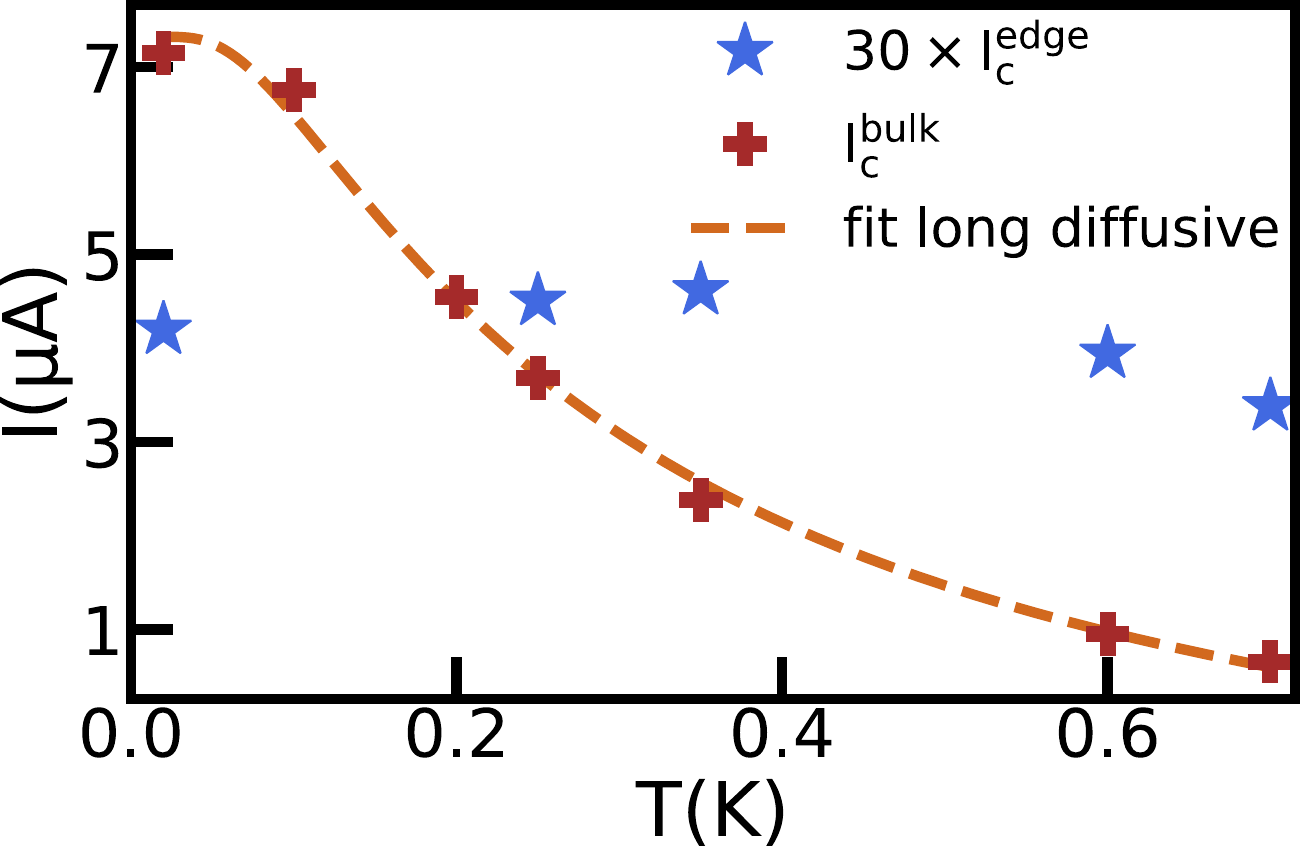}
	\caption{Temperature dependence of the critical current of SQUID C ($I_c^{bulk}$), compared the edge junction's CPR $I^{edge}$ (multiplied by 30 for visibility). The amplitude of the critical current for the temperature 250 mK, 350 mK, 600 mK and 700 mK is obtained by reconstructing the signal from all the harmonics extracted from Fig. \ref{fig:Fig_2} and then taking the maximum of this curve. The shape of the CPR being dominated by the first harmonic, the dependence is extremely close to the one of the first harmonic alone (Fig. \ref{fig:Fig_suppmat_shortorlong}). The greater resilience to temperature of the edge supercurrent is clear.}
	\label{fig:Fig_suppmat_IcT}
\end{figure}

\subsection{Contrasting the effect of an imperfect NS interface on the CPR of a ballistic junction in the topological versus non topological case}

In this section we illustrate how disorder affects differently the CPR of a ballistic junction depending on whether or not topological protection causes the ballistic behavior. As demonstrated in \cite{cayao_andreev_2018,murani_andreev_2017}, disorder at the NS interface affects differently the harmonics content of the CPR for a trivial or a topological junction. Because backscattering by non-magnetic impurities in the topological wire is forbidden  \cite{adroguer_probing_2010}, topological (i.e. spin-momentum-locked, helical) channels are expected to exhibit a perfect Andreev reflection  at the NS interface  even if the interface is not perfectly transparent,  $t_{NS}<1$. One therefore expects a robust discontinuity at phase $\pi$ in the CPR at zero temperature (i.e. same as for $t_{NS}=1$)  whose amplitude however decreases as $t_{NS}$ decreases. This contrasts with the case of ballistic trivial (non-topological) junctions, whose harmonics content is very sensitive to the NS interface transmission: in such junctions, the discontinuity at $\pi$ gets rounded by an imperfect interface, and the amplitude of the critical current decreases faster with $t_{NS}^2$. Figs. \ref{fig:Fig_suppmat_tNS_nontopo} and \ref{fig:Fig_suppmat_tNS_topo} compare the effect of decreasing NS interface transmission on two ballistic SNS junctions, a topologically protected one and a non-protected one. It is clear that the sawtooth shape is preserved  in the case of the topologically protected junction, but not in the other junction, in which rounding of the CPR appears as soon as $t_{NS}$ is not 1.

\section{Appendix B: Rounding caused by the asymmetric SQUID configuration}

As pointed out by \cite{babich2023,endres2023,alexandrebernard2022}, a CPR with a discontinuity will appear rounded in an asymmetric SQUID measurement, because the phase of the reference junction adjusts to avoid the discontinuous decrease in the supercurrent of the weak junction. This causes the SQUID's critical current to reflect the CPR of the reference junction over a (small) portion of the flux variation, as illustrated in Fig. \ref{fig:Fig_suppmat_sinsaw}.  Because this effect is smaller than the intrinsic CPR rounding given by the t=0.9 coefficient, also plotted in the figure, even at zero temperature, we can safely attribute the rounding in the experiment to the edge junction's intrinsic small rounding of the sawtooth CPR. 

\section{Appendix C: Additional data and information on data processing for SQUID C}
\subsection{Extraction of the Current-phase relation}
The current-phase relation is obtained from the measurement of $\frac{dV}{dI}(B,I_{dc})$ as a function of magnetic field $B$ and DC bias current $I_{dc}$, measured using a standard lock-in measurement technique. One obtains typically curves like the one on Fig. \ref{fig:Fig_suppmat1}.

An example of the extracted critical current $I_c$ is plotted in red in Fig. \ref{fig:Fig_suppmat2}.

We note a feature in the dV/dI measurement around $I_{dc}\approx \SI{-1.3}{\micro\ampere}$. We attribute it to the switching to the normal state of a junction in series with the measured SQUID. This junction is most likely formed between this SQUID and a second, nearby SQUID, which happened to be connected to a line grounded half way up the dilution refrigerator. This causes an unbalance in the thermoelectric voltages of the in and out measurement lines, generating a net voltage, and explaining why the feature is not centered around $I_{dc}=0$.

We also find a small offset current that is mostly visible in the high field data.  This current may be due to a supercurrent along edges of the junction outside the junction area. Its magnitude can be estimated to vary between 100 and 200 nA depending on temperature, by checking for time-reversal symmetry, see the paragraph on the superconducting diode effect. We have therefore shifted the zero current line in the main text's Fig. \ref{fig:Fig_3}(b), as indicated in the caption. 

To treat the dV/dI data, we first subtract the background resistance $R_{min}\approx 172~ \Omega$, corresponding to the wiring resistance (these are two wire measurements). We then define $R_N=R_{max}-R_{min}$ the normal state resistance. The SQUID's critical current is then extracted as the bias current at which the differential resistance reaches a threshold defined as $R_{T}=R_N/2$. If the threshold falls between two discrete measurement points, we interpolate $I_c$ as described in the inset of Fig. \ref{fig:Fig_suppmat1} to reduce discretization.

The current-phase relation is visible on top of a modulation on a large (several mT) field scale, originating from the Fraunhofer-like pattern due to interference across the reference junction. 
The critical current versus field curve is numerically low-pass filtered to remove this contribution. To this end, the Fourier transform is computed over the full data range ($B_{tot}=8~mT$,  which is slightly larger than three periods of the current-phase relation) and the three first frequencies are cut ($f=0$, $f=\frac{1}{B_{tot}}$ and $f=\frac{2}{B_{tot}}$). The CPR is then computed as the inverse Fourier transform, yielding the data presented in the main text, and in blue in Fig. \ref{fig:Fig_suppmat2}.

\subsection{Fourier transform analysis}
To analyze the harmonic content of the CPR, we compute its Fourier transform. To this end, we prepare the data by selecting an integer number of periods (here three) and interpolating the data in order to get a table of $2^8=256$ points that exactly match the three periods. With data acquired on such a small number of periods, this operation is crucial to visualize properly the lowest and higher harmonics (frequencies). Indeed, we recall that the resolution of the calculated Fourier transform is given by $\frac{1}{B_{tot}}=\frac{1}{3B_0}$ with $B_{tot}$ the total measurement range and $B_0$ the lowest B-period, implying that the amplitude is calculated for all multiples of $\frac{1}{3B_0}$. Consequently, if the only frequencies present in the signal are $B_0^{-1}$ and its harmonics, they should be captured by the calculation.

We then compute the absolute value of the FFT calculation using the fft.rfft function of the numpy library in python, without windowing (windowing would be useful if we didn't select an integer number of periods), and renormalize by multiplying by $2/N$ ($N$ the number of points of the input data).
Finally, we numerically filter out the lowest two frequencies, namely the peaks at $0$ and $\frac{1}{3B_0}$ are set to zero. This leads to the curves presented in Fig. 3.

We estimate that the uncertainty on the harmonics' amplitude is of the order of the noise at high frequency, corresponding to $3~nA$. This is an uncertainty at 95\%, meaning that there is 95\% chance that the "real" value is in the interval $[x_{mes}-dx,x_{mes}+dx]$.

The next step is to fit the harmonics' amplitude by the function $\tilde I(n)=\tilde I_0\frac{e^{-an}}{n}$, with 
$a = 2\pi k_BT/E^{edge}_{Th}-\ln(t)$.
To get rid of the parameter $I_0$, we instead fit $\frac{\tilde I_n}{\tilde I_1}=\frac{e^{-a(n-1)}}{n}$, with $a$ the only fitting parameter. Fig. \ref{fig:Fig_suppmata} displays the extracted values of $a$ for four temperatures, and the fit to a linear temperature dependence, from which we extract $t=0.89\pm0.04$ and $E^{edge}_{Th}=560\pm75\mathrm{~\mu eV}\simeq ~6.5~K$.
To determine the uncertainty on the parameter $a$, given the uncertainty on $\tilde I(n)$, we use a Monte Carlo method. The idea is to generate numerically a large number (here $N=1000$) of new data sets that are consistent with uncertainty. Each of these data set is fitted and provides a value of $a$. An estimate of the uncertainty at 95\% on $a$ is given by twice the standard deviation of the $N$ values of $a$ obtained. The uncertainties on $t$ and $E_{Th}^{edge}$ are also estimated using the Monte Carlo method.

\subsection{Short versus long ballistic junction; estimate of the superconducting coherence length in the ballistic regime using an estimate for the superconducting gap at the interface}
We have seen that the CPR of a ballistic junction takes the shape of a sawtooth, i.e. the characteristic "long junction" CPR, as soon as the junction is more than a few atomic sites long. The difference between long and short ballistic regime is also hard to determine from the decay with temperature of the critical current in both regimes, since the decay has a similar close-to-exponential-like dependence, with a characteristic energy given by either the Thouless energy or the contacts' superconducting gap.
To identify the junction regime, we therefore estimate the superconducting coherence length using the junction parameters, and determine whether the junction length is smaller (short junction regime) or  greater (long junction regime) than this coherence length. 
The superconducting coherence length in a ballistic junction is $\xi=\frac{\hbar v_F}{\pi \Delta}\approx 200-600 \mathrm{~nm}$ 
(using the helical state Fermi velocity $v_F=5\times 10^5\mathrm{~m.s^{-1}}$ estimated in \cite{Fei2016}) and $\Delta \approx 170-500~\rm{\mu eV}$. The lower estimate for $\Delta$ is the value reported in \cite{Delagrange2015,Delagrange2016} for a Pd(7 nm)/Nb(20 nm)/Al(40 nm), AlOx and Al(120 nm) multilayer contacting carbon nanotubes. The gap in the present experiment's Pd (8 nm)/Nb(80 nm) bilayer contact may be higher because of the thickner Nb, and the fact that Pd has been shown to form a superconducting compound with Te.  We therefore consider that the 600 nm-long edge junction is in the intermediate regime between short and long. In such a regime the CPR is expected to have a sawtooth shape.

\subsection{Bulk junction}
In this section we give a more detailed view of the bulk junction, argue that due to its geometry it must act as the reference junction, and show that this hourglass-shaped junction on the bulk of the SQUID C's surface behaves equivalently to a rectangular-shape junction, whose parameters we estimate. 

\paragraph{Details of the bulk junction geometry}
The bulk junction has the form of an hourglass, with two triangular superconducting electrodes with a small separation of less than 100 nm, as can be seen from the Scanning Electron Micrographs of Fig. \ref{fig:SEMbowtie}.  The hourglass junction layout is sketched in Fig. \ref{fig:triangle}, defining the junction's parameters $y_{max} = 0.73$ µm, $t = 223$ nm the thickness of the flake, and $d = 2.21$ µm the length of the diagonal of the hourglass (used to calculate the angle), $l(y_{max}) = 2.06$ µm. 

\paragraph{Why this junction acts as the reference junction in the asymmetric SQUID}

The hourglass geometry, with its very closely spaced superconducting electrodes, was specifically chosen to create a reference junction with as small a surface as possible. This ensures that the junction's critical current is quite large, and also enables to follow the magnetic field dependence up to the high fields (indeed a junction’s critical current dies out at fields that vary inversely proportionally to the junction area). 
We would like to insist on the fact that such an unconventional shape is by no means a hindrance to this junction acting as a reference junction, since the only criterion for a junction in an asymmetric SQUID to play the role of the reference junction is that its critical current be much larger than the critical current of the other junction (whose CPR is sought).
Note that it is possible that diffusion of Pd from the electrodes into WTe2 could occur and generate a superconducting Pd1-xTex compound that would connect the two electrodes. We believe that even if the weak link in the bulk junction is a PdTex superconducting filament rather than (or in parallel with) the proximitized bulk WTe2, that would be a valid weak link with which to measure the CPR of the edge junction. 
In the present case, as we present in Fig.  \ref{fig:comparison_edge_bulk_junctions}, it is clear that the critical current of the bulk junction must be much larger than that of the edge junction: the edge junction’s superconducting electrodes are 600 nm apart, whereas the bulk junction’s electrodes are much closer (as seen in the SEM image), and may also have a parallel conduction via the PdTex filament. Consequently, the small (150 nA) oscillation must be due to the junction with the electrodes furthest apart, i.e. the edge junction.

\paragraph{Equivalent model for the hourglass-shaped junction}
In this section we show that the hourglass-shaped junction on the bulk of the SQUID C's surface behaves equivalently to a rectangular-shape junction, whose parameters we estimate.  To this end, we evaluate the junction's resistance and find the dimensions of the rectangular junction with the same resistance. 

We first point out that the change in resistance above the SQUID's critical current is the parallel combination of the bulk and edge junctions resistances, but we can check a posteriori that the bulk junction resistance is smaller than the edge junction resistance and therefore that the change in resistance is practically equal to the bulk junction resistance.  This resistance is roughly $R_N\simeq 1 \Omega$, see the $\mathrm{dV/dI(I_{DC})}$ curve in Fig. \ref{fig:dVdI_single}.

The hourglass junction, sketched in Fig. \ref{fig:triangle}, is characterized by the following parameters : $y_{max} = 0.73$ µm, $t = 223$ nm the thickness of the flake, and $d = 2.21$ µm the length of the diagonal of the hourglass (used to calculate the angle), $l(y_{max}) = 2.06$ µm. 
We assume that the junction is symmetric and for the calculations we thus only consider the top, triangular-shaped half of the hourglass, starting at $y = 0$ up to $y = y_{max}$. We define an elementary conductance $g(y)$ as :

\begin{align}
    g(y) = \sigma \frac{S}{2l(y)+l_0}
\end{align}
with $S = t\times dy$, $\sigma $ the conductivity and $l_e \simeq 50$ nm the mean free path. The angle $\alpha$ is defined as :
\begin{align}
    \tan(\alpha) = \frac{y}{l(y)}=\frac{y_{max}}{l(y_{max})}=0.35.
\end{align}
This allows to write the elementary conductance as
\begin{align}
    g(y) = \frac{\sigma t dy}{2\frac{y}{\tan(\alpha)}+l_e}
\end{align}
so that the conductance of the junction is :
\begin{align}
    \begin{split}
        G &= 2\times \int_0^{y_{max}} \frac{\sigma t}{2 y/\tan(\alpha)+l_e} dy\\
        &= \sigma t \tan(\alpha) \ln(\frac{2y_{max}}{\tan(\alpha)l_e}+1)
    \end{split}
    \label{eq:ConductanceTriangle}
\end{align}
Using $\sigma = 2~10^6 \mathrm{~\Omega^{-1} m^{-1}}$  \cite{choi2020}, we obtain :
\begin{align}
    R = \frac{1}{G} = 1 - 1.5~ \mathrm{\Omega}, 
\end{align}
close to the value found in the experiment for both bulk and edge junctions in parallel, confirming that the bulk junction's resistance is smaller than the edge junction. 
With the expression of $G$ obtained in equation \ref{eq:ConductanceTriangle}, we can compute an effective length for the hourglass junction using
\begin{align}
    G = \frac{\sigma S}{L_{\rm{eff}}},
\end{align}
with
\begin{align}
    \begin{split}
        L_{\rm{eff}} &= \frac{2y_{max}}{\ln(\frac{2y_{max}}{l_0\tan(\alpha) }+ 1)}\times\frac{1}{\tan(\alpha)}\\
        &= 0.94 \text{  \textmu m}
    \end{split}
\end{align}

and the hourglass junction is equivalent, as a first approximation, to a rectangular junction of width $w = 2y_{max}$ and length $L_{\rm{eff}}$, as illustrated in Fig. \ref{fig:transformation_hourglass}.

In the next section we show that the Thouless energy describing the bulk junction's experimental behavior agrees with the Thouless energy computed for the equivalent rectangular junction. 

\subsection{Extraction of the Thouless energy $E^{bulk}_{Th}$ from the temperature dependence of the critical current of SQUID C's bulk junction, and comparison with $E^\mathrm{bulk,eff}_{Th}$ of the equivalent rectangular junction}

In the main text, we use an approximate expression for the temperature dependence of a long diffusive junction,  $I^{bulk}_{c}(T) \simeq \exp(-\frac{T}{T_c})$, with a characteristic temperature $T_c$ related to the Thouless energy via $T_c=3.8 E^{bulk}_{Th}$ \cite{Li2016,Wilhelm}. This yields $T_c\simeq 250 \mathrm{~mK}$, and $E^{bulk}_{Th}\simeq 65~\mathrm{ mK} \approx 6~ \mathrm{\mu eV}$, much less than the superconducting gap, which is greater than $2 \mathrm{~K},$ confirming the disordered long junction regime.
In fact, there are more appropriate expressions that should be used to deduce $E^{bulk}_{Th}$ from  $I^{bulk}_{c}(T)$, as described in the following.

The critical current extracted from the differential resistance curves  such as presented in Fig. \ref{fig:dVdI_single} is plotted as a function of temperature in Fig. \ref{fig:Fig_suppmat_IcT}. Extracting the Thouless energy  from the temperature dependence relies on fitting to the appropriate expression, which depends on the junction and temperature regime:

\textbf{Long diffusive junction}

The long diffusive regime corresponds to $\Delta>100E_{\rm{Th}}$. Depending on the temperature regime, two different expressions should be used, as described in \cite{PhDDubos}: At low temperature ($T<5E_{\rm{Th}}$), the decrease of the critical current is fit with:
\begin{equation}
   eR_NI_c(T)=10.82 E_{\rm{Th}} (1-1.3\mathrm{e}^{-\frac{10.82 E_{\rm{Th}}}{3.2k_BT}}),
\end{equation} 
whereas at higher temperature the expression is:
\begin{equation}
    \begin{split}
       eR_NI_c(T)= & \frac{32}{3+2\sqrt{2}} E_{\rm{Th}}(\frac{2\pi k_B T}{E_{\rm{Th}}})^{\frac{3}{2}} \\
        &\times \sum_{n=0}^{n=\infty} \sqrt{2n+1}\mathrm{e}^{-\sqrt{2n+1}\sqrt{\frac{2\pi k_B T}{E_{\rm{Th}}}}}.
    \end{split}
\end{equation} 
We find that the entire data set for SQUID C's bulk junction, from 10 mK to 700 mK, can be fit with both expressions, and yields the same Thouless energy, $E_{\rm{Th}} = 72$ $\mathrm{mK}$, with $5E_{\rm{Th}}\simeq 350~mK$.

\textbf{Long to short intermediate behavior}
Since $\Delta/E_{\rm{Th}}\simeq30<100$,  the junction is not fully in the long diffusive regime, and the decrease in temperature of the critical current is governed by a different decay factor. The general low-temperature behavior is given by an expression similar to the long junction expression, but with a factor $b$ replacing the factor 10.82 (\cite{PhDDubos}):
\begin{equation}
   eR_NI_c(T)=b E_{\rm{Th}} (1-1.3\mathrm{e}^{-\frac{b E_{\rm{Th}}}{3.2k_BT}}).
\end{equation} 
 $b$ depends on $\frac{\Delta}{E_{\rm{Th}}}$, $b=10.82$ for a very long junction but $b=7.7$ for $\frac{\Delta}{E_{\rm{Th}}}=20$. The previous fits are unchanged but now the value of $E_{\rm{Th}}$ is $\frac{10.82}{7.7}=1.4$ times larger. These considerations therefore give an estimate of the uncertainty  on the values of $E_{\rm{Th}}$ determined. 

The result for SQUID C's bulk junction   $E_{\rm{Th}}^{\rm{bulk}} = 72$ $\mathrm{mK}$ is consistent with an estimate of the Thouless energy of the rectangular junction that approximates the bow-tie bulk junction:
\begin{align}
    \begin{split}
      E^{\mathrm{bulk,eff}}_{\rm{Th}} &= \frac{\hbar D}{L_{\rm{eff}}^2}\\
        &= \frac{\hbar v_F l_e}{3L_{\rm{eff}^2}}\\
        &\simeq 30~ \text{  mK},
    \end{split}
\end{align}
using $L_\mathrm{eff}\simeq 0.94~\mu m$, $l_e\simeq 50~nm$ (\cite{Zhang2021}) and $v_F\simeq2\times10^5 m/s$, a Fermi velocity typical of the bulk electron and hole pockets, averaged over a,b, and c directions \cite{li2017a}. 

Another indication that the rectangular junction may be a good approximation of the bow-tie shaped junction is the Fraunhofer-like pattern observed for $I_c^{\rm{bulk}}$(B), which is the pattern expected of a wide rectangular-shaped junction \cite{Chiodi2012,Cuevas2007}.

These considerations confirm the diffusive nature of transport in the hourglass-shaped bulk junction and the validity of approximating the bow-tie junction with an equivalent rectangular junction. 
Of course, these estimates suffer from the fact that the expressions considered are those for perfect NS interfaces, which is not the case in the experiment, as discussed in the main text.

\subsection{Illustration of the very different energy scales characterizing the edge and bulk junctions of SQUID C}

Figure \ref{fig:Fig_suppmat_IcT} plots on the same graph the temperature dependence of the critical current of SQUID C's bulk junction at zero field, $I_c^{bulk}$, and that of the edge junction's critical current $I_c^{edge}$ (multiplied by 30 for visibility). The greater resilience to temperature of the edge critical current is clear, and translated into two very different Thouless energies, $E^{bulk}_{Th}\simeq 70~mK$ and $E^{edge}_{Th}\simeq  6.5~K$ or $E^{edge}_{Th}\simeq  2.8~K$, depending on whether $E^{edge}_{Th}$ is estimated using the decay of the CPR's harmonics (as in the main text and Fig. \ref{fig:Fig_suppmata}) or by fitting the temperature dependence of the CPR's first harmonic, as in Fig. \ref{fig:Fig_suppmat_shortorlong}. 
    
\subsection{Superconducting Diode effect in SQUID C}

\begin{figure}[tb]
	\centering
	\includegraphics[width=\columnwidth]{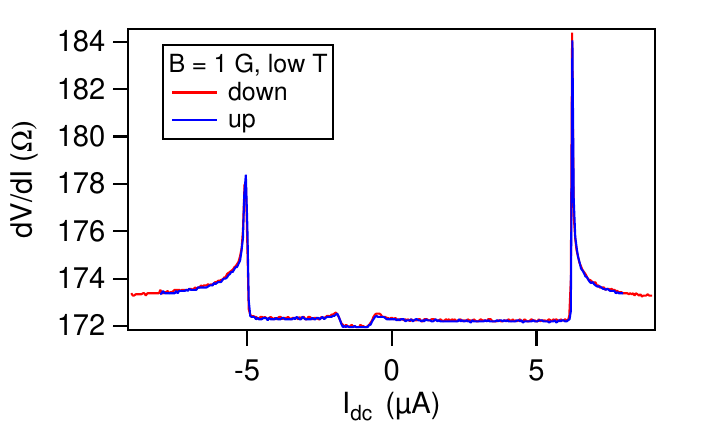}
	\caption{Differential conductance at the fonction of DC bias current at $B\approx 1 G$ and at the lowest temperature, around 20 mK. The critical current is extracted as the current at which the differential resistance jump exceeds a threshhold of a fraction of the normal state resistance. }
	\label{fig:Fig_suppmat3}
\end{figure}

To examine the superconducting diode-like behaviour, we compare the magnetic field dependence of the critical current measured for positive current, $I_{c,+}(B)$, with the one measured with negative current $I_{c,-}(B)$. Specifically, $I_{c,+}$ is the switching current measured when the current is ramped up from the superconducting state and $I_{c,-}$ the absolute value of the switching current when the current is ramped down from the superconducting state.
Measuring both quantities as a function of $B$ would be quite time consuming. Instead, we start by comparing a measurement of the differential conductance as a function of bias current at a given magnetic field for both ramping up and down of the current, represented on Fig. \ref{fig:Fig_suppmat3}.

\begin{figure}[tb]
	\centering
	\includegraphics[width=\columnwidth]{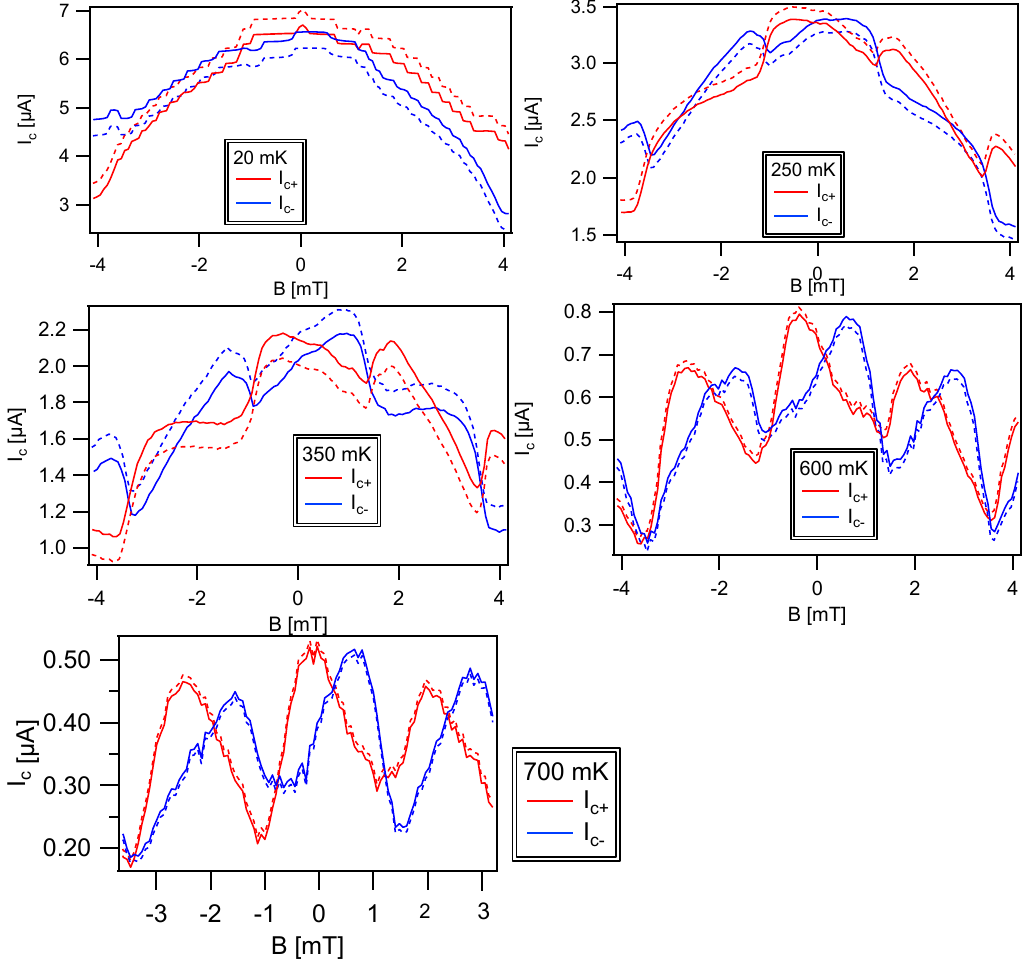}
	\caption{ SQUIDs  positive ($I_{c+}$, red) and negative  ($I_{c-}$, blue) critical current as a function of magnetic field B for the five measured temperatures. If time-reversal symmetry is preserved, one should have $I_{c+}(B)=I_{c-}(-B)$. This equality is roughly (though not exactly) verified if one corrects for the small offset current (between 100 and 200 nA, depending on temperature) mentioned in the text, see dashed lines.
 }
	\label{fig:Fig_Diode}
\end{figure}

Surprisingly, we find that both curves are exactly the same. This was not a priori expected, since the retrapping current (at which the junction switches from normal to superconducting when decreasing the current) can be very different from the switching current (at which the junction switches from superconducting to normal) if the junction is in an underdamped regime or in the presence of heating effects. Here, the retrapping and the switching are exactly the same (and we checked that this was the case in different experimental conditions), corresponding to an overdamped regime. This is actually very convenient, since it implies that a single up-sweep measurement of $dV/dI(I_{dc})$ yields both $I_{c,+}$ (given by the switching current) and $I_{c,-}$ (given by the retrapping current). We can then compare $I_{c,+}$ and $I_{c,-}$ for all temperatures, see Fig. \ref{fig:Fig_Diode}.

The large diode effect found here is not surprising, since it is expected of an asymmetric SQUID in a magnetic field in which the weak link has a non-sinusoidal CPR.

\subsection{SQUID C's critical current at two different temperatures}

\begin{figure}[tb]
    \centering
    \includegraphics[width=\columnwidth]{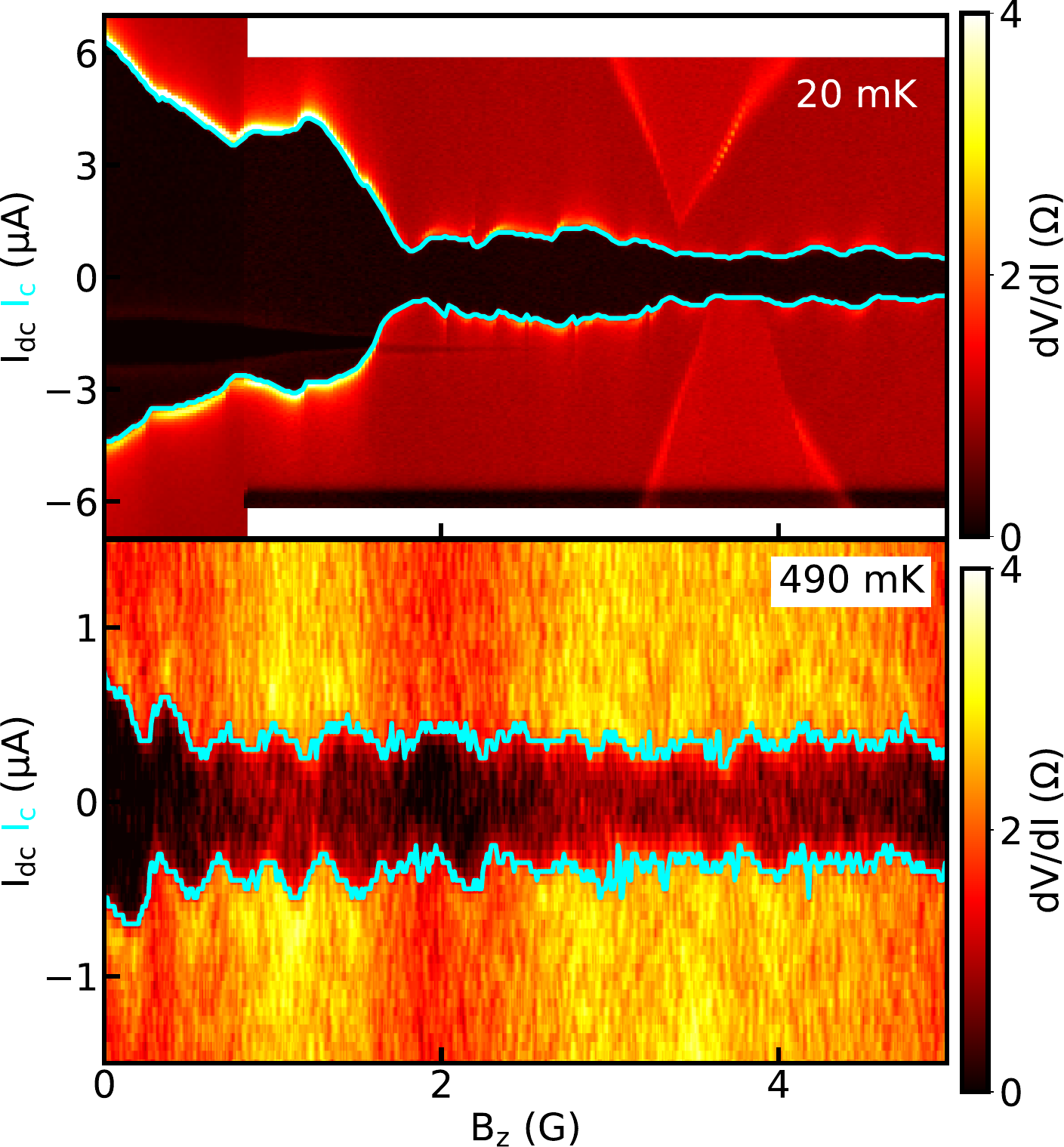}
    \caption{Differential conductance as a function of current and perpendicular field Bz for SQUID C, at two temperature. Top: T=20 mK and bottom, T=490 mK.  Bz is 2$\%$ of the applied magnetic field, as explained in the text. The extracted critical current is plotted in cyan. The ac current is between 100 and 20 nA depending on the field range for the 20 mK plot and is 5 nA for the 490 mK plot.}
    \label{fig:SM_20mK490mK}
\end{figure}

\begin{figure}[tb]
    \centering
    \includegraphics[width = \columnwidth]{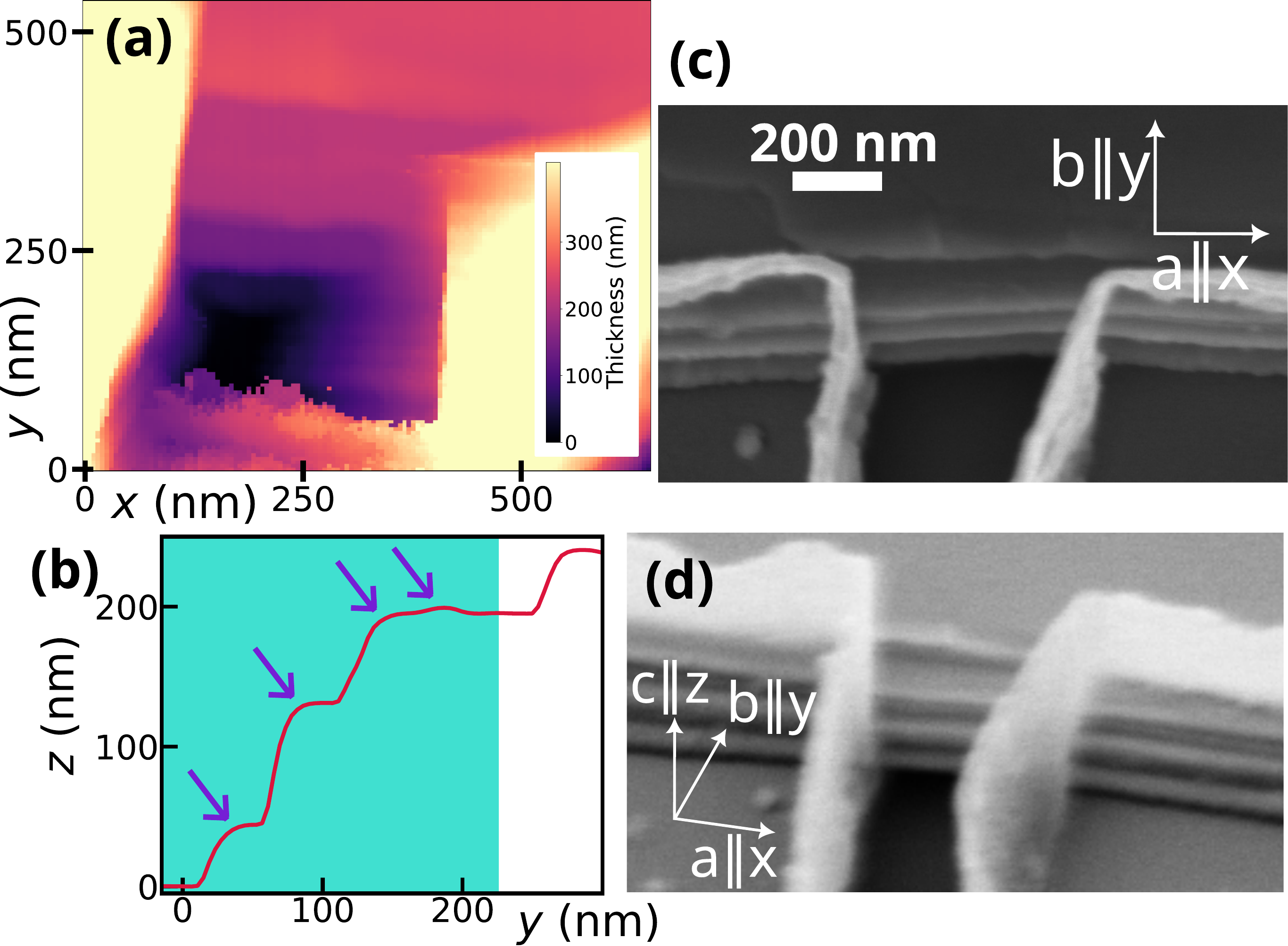}
    \caption{Edge junction of SQUID C, as seen with Atomic force Microscopy and Scanning Electron Microscopy. (a) AFM topographic image of  the junction area, with a line cut  along the b axis, showing the giving the height profile of the junction. (c) and (d), SEM images at a small and larger tilt angle so that the topography is clearer in (d). Five steps are visible, with height ranging from less than 10 nm to 90 nm, and terrace width varying between 20 and 100 nm. Only the first four steps are covered by the superconducting contacts (hatched in blue). }
    \label{fig:AFMSQUIDC}
\end{figure}

Fig. \ref{fig:SM_20mK490mK} compares SQUID C’s field dependence between 0 and 25 mT in B, i.e. 0 and  5 G in Bz, at two temperatures: the 20 mK plot, which is over a smaller range than shown in Fig. \ref{fig:Fig_3}, and a 490 mK plot. Both plots display more than ten clear oscillations with the period of 0.46 G corresponding to the SQUID geometry. 

\subsection{Calculated interference patterns for different configurations of 1D channels in the edge junction of SQUID C}
 Although it is probably vain to hope to extract from the measured interference pattern the exact number and spatial distribution of 1D states in the edge junction of SQUID C, we can nonetheless simulate the interference patterns corresponding to  slightly different numbers and distributions of hinge states, taking as an input the actual layout of the edge junction of SQUID C, as pictured in Fig. 26.
 
We simulate the expected critical current versus magnetic field with the Temperature dependence of the critical current of SQUID C (see Fig. 24). 
The model consists of:
\begin{itemize}
    \item a reference bulk junction with a critical current  $I_{c,bulk}=\mathrm{7~\mu A}$ and a Fraunhofer (sinc) pattern with 6 Gauss-size side lobes;
    \item a SQUID loop area  $S_{\rm{SQUID}}=\mathrm{12\mu m\times1.2\mu m}$; 
    \item ballistic channels on the edge that do not decay in magnetic field.
\end{itemize}
The total critical current is computed as:
\begin{equation}
 \begin{split}
    I_c(B)&= \max_\varphi~ [I_{c,\rm{bulk}} f_{\rm{diff}}(\frac{BS_{\rm{SQUID}}}{\Phi_0}+\varphi)\\
    &+\sum_{i=1}^{N_{channels}} I_{c,i} f_{\rm{ball}}(\frac{BS_i}{\Phi_0}+\varphi)],
    \end{split}
\end{equation}
where $S_i$ is the area between channel i and channel 1.
The critical current depends on the spatial separation between the different channels. We consider several cases, for which the total critical current, $\mathrm{210~ nA}$ is evenly split between the different channels:
 \begin{enumerate}
    \item one channel with a  $\mathrm{210~ nA}$ critical current;
    \item two channels, each with a $\mathrm{105~ nA}$ critical current, whose separation defines an area that is 2\% of the SQUID area;
    \item three channels, each with a  $\mathrm{70~ nA}$ critical current, regularly spaced, defining an area that is 2\% of the SQUID area;
    \item three channels, each with a  $\mathrm{70~ nA}$ critical current, whose separation defines an area that is respectively 2\% and 3\% of the SQUID area. 
    \item ten channels of critical current $\mathrm{21 ~nA}$ each, separated by a random fraction of the SQUID area between 1\% and 3\%.    
\end{enumerate}

Figure 24 
displays the interference patterns computed around zero-field, at low field and high field  for the different edge state distributions. (a) The sawtooth pattern appears as a modulation in all cases in the first lobe. (b) At intermediate magnetic field, the sawtooth modulation splits into several smaller sawtooths if there are several ballistic channels, blurring the current-phase relation. (c) At high magnetic field, the current in the diffusive bulk junction has become smaller than the supercurrent through the edge junction, so the modulation is due to the (sinus-like) bulk junction's CPR. The envelope corresponds to the sum of the ballistic channels and depends on their number and separation. If the channels are regularly spaced, we see a periodic pattern with a period related to the surface area separating two of those ballistic channels. However, if there are more than two channels with different separations, the pattern will look erratic on the same field scale. The greater the distance between two channels, the faster they contribute differently to the critical current and the faster the periodic pattern becomes blurred. A higher number of irregularly positioned channels leads to an average critical current given by the square root of the number of channels times the critical current of one channel. (d) For the simulation on SQUID C, we use as the separation between channels the distance between steps that we measure with Atomic Force Microscopy. We compute the interference pattern with three such channels (shown with blue circles in panel (d)), but we do not claim that those are the actual ones that carry the supercurrent in the experiment. In the present experiment the magnetic field is mostly in the sample plane, so the vertical distance between steps may be more relevant that the horizontal one. (e) SQUID with a diffusive bulk junction and an edge junction carrying 3 ballistic channels. We assume that no diffusive current runs through the edge junction. (f) and (g) superimpose the simulated critical current on top of experimental data.

\begin{figure*}[tb]
	\centering
    \includegraphics[width=0.9 \textwidth]{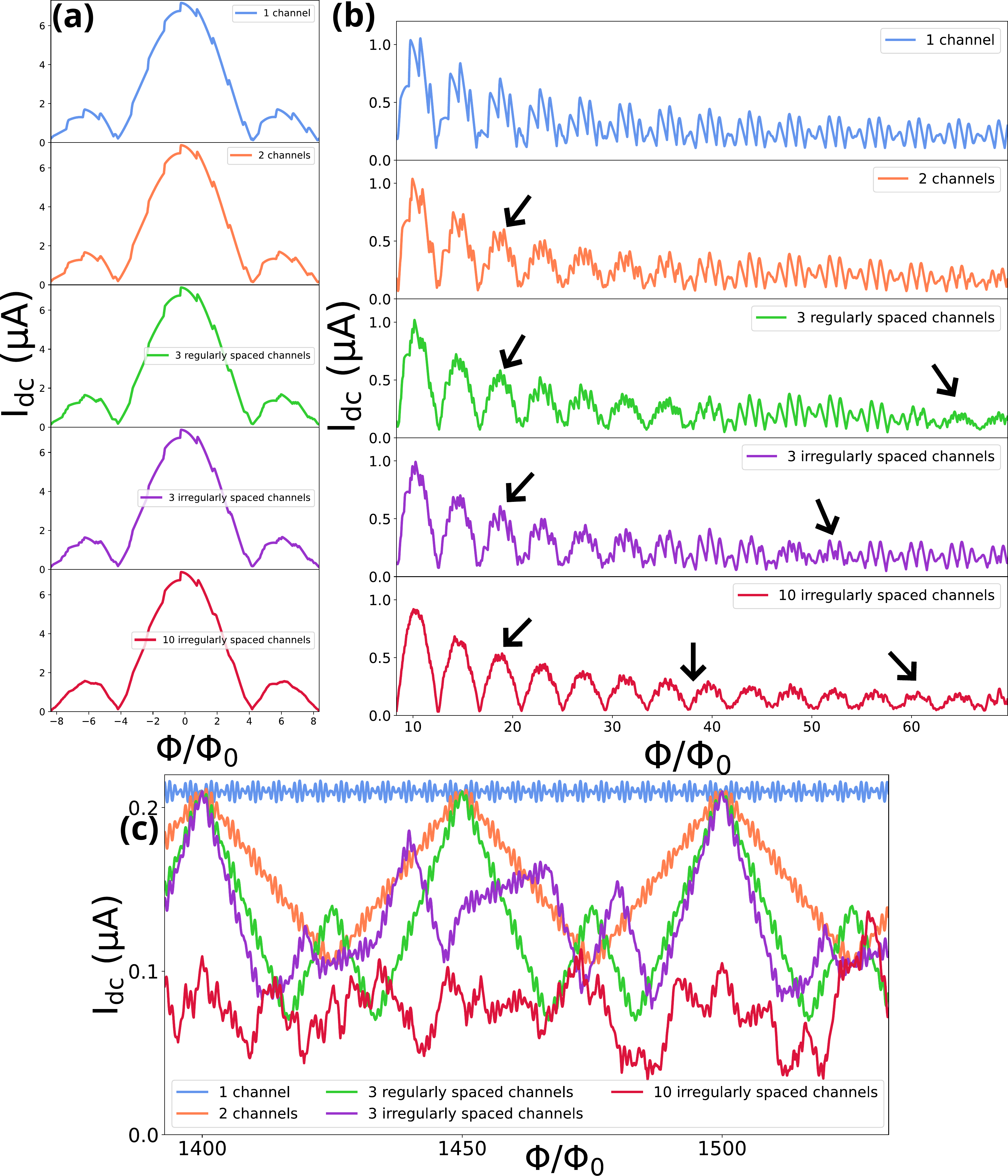}
	\caption{Simulation of interference patterns for different number and configurations of ballistic channels in the edge junction, assuming that the reference bulk junction is wide and decays with a fast Fraunhofer pattern. (a) The sawtooth pattern appears as a modulation in all cases in the first lobe. (b) At intermediate magnetic field, the sawtooth modulation splits into several smaller sawtooths if there are several ballistic channels, blurring the current-phase relation. Black arrows show such areas up to tens of $\frac{\Phi}{\Phi_0}$. The magnetic flux at which it happens depends on the separation betwee ballistic channels. (c) At very high magnetic field, the current in the diffusive bulk junction has become smaller than the supercurrent through the edge junction, so the modulation of period $\Phi_0$ is due to the (sinus-like) bulk junction's CPR. The envelope corresponds to the sum of the ballistic channels and depends on their number and separation, it is triangular for two channels (orange on the figure). Regularly spaced channels lead to a periodic pattern with a period related to the surface area separating two of those ballistic channels. However, if there are more than two channels with different separations, the pattern will look erratic on the same field scale. A higher number of irregularly positioned channels leads to an average critical current given by the square root of the number of channels times the critical current of one channel.}
    \label{fig:SM_channels}
\end{figure*}

\begin{figure*}[tb]
	\centering
    \includegraphics[width=0.9 \textwidth]{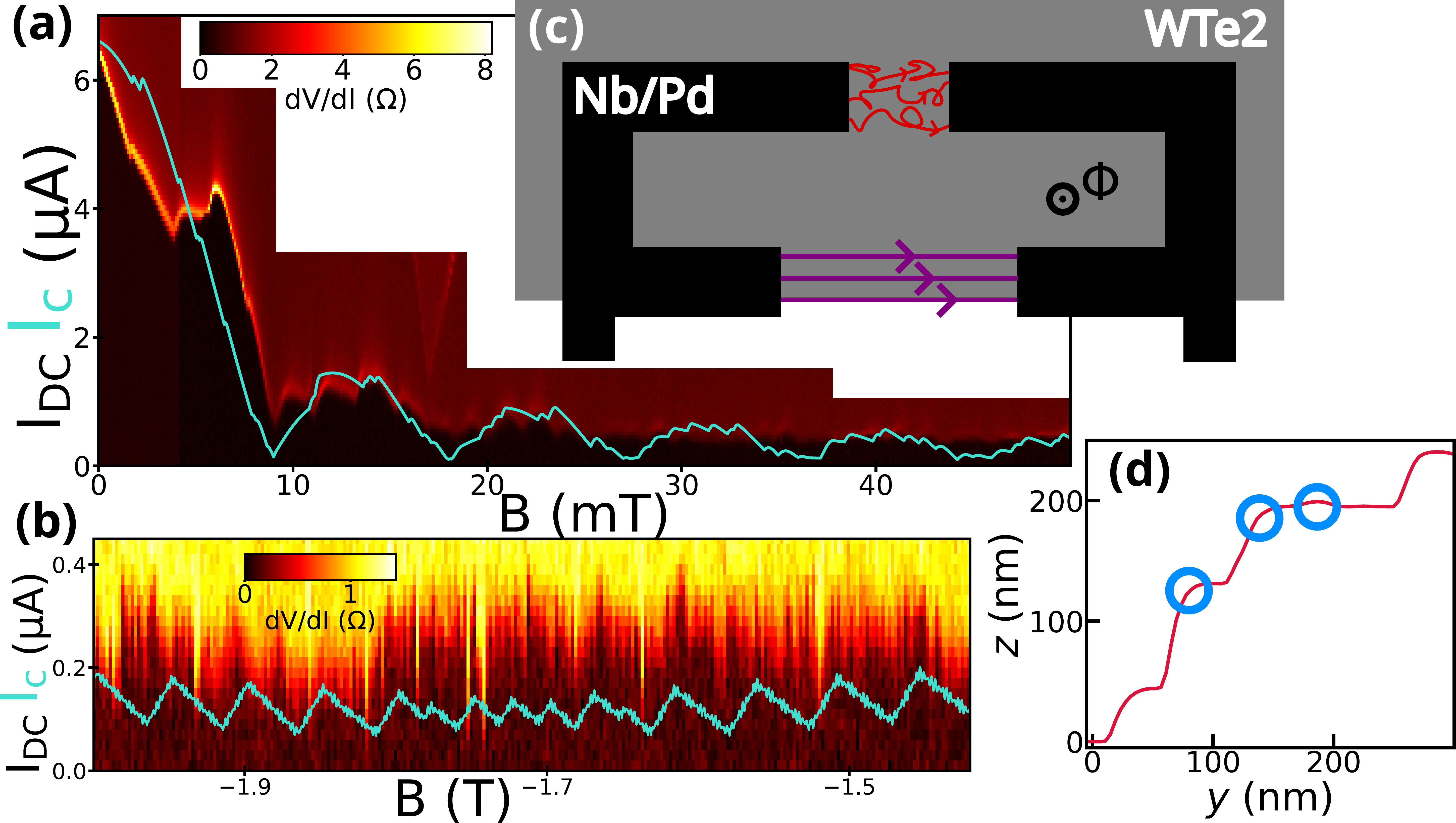}
       \caption{(a) and (b) Simulated critical current (cyan) layed over  experimental data of SQUID C.(c) Sketch of a SQUID with a diffusive bulk junction and an edge junction with three ballistic channels. We suppose that no diffusive current runs through the edge junction. Blue line in (a) and (b): Calculated critical current of a SQUID in which one junction has a sinusoidal CPR with a zero field critical current of $6.5~\mu A$, decaying as a Fraunhofer (sinc) function, and the other junction is made of three 1D channels, each with a 70 nA sawtooth CPR, that decays on a 10 T scale. (d) For the simulation on SQUID C, we use as the separation between channels the distance between steps measured with Atomic Force Microscopy. The interference pattern shown in cyan in (a) and (b) was computed with three channels positioned as shown by the blue circles in (d), but we do not claim that those are the actual ones that carry the supercurrent in the experiment.   }
       
       \label{fig:SM_SQUIDC_channels}
\end{figure*}

\section{Appendix D: Comparison of all SQUIDs}
\begin{table*}
\begin{tabular}{|c|c|c|c|c|}
  \hline
  \multicolumn{2}{|c|}{SQUID} & $E_{\rm{Th}}$ (mK) & $D=\frac{L^2}{\hbar}E_{\rm{Th}}$ (m$^2.$s$^{-1}$) & $\frac{10.82E_{\rm{Th}}}{eR_NI_C}$ \\
  \hline
  C & $I_{bulk}$ (from CPR measurement) & 72 & 0.0084 & 10 \\
  \hline
  C & $I_{edge}$ (see main text) & 2-6 K & & \\
  \hline
  A (2022-10) & $I_c$ & 34 & 0.0011 & 16 \\
  \hline
  \multirow{2}{*}{B1 (2023-01)} & $I_{bulk}$ (from CPR measurement) & 51 & 0.0024 & 16 \\
  \cline{2-5}
  & $I_{edge}$ (from CPR measurement) & 39 & 0.0051 &  \\
  \hline
  \multirow{2}{*}{B2 (2023-01)} & $I_{bulk}$ (from CPR measurement) & 78 & 0.0037 & 36 \\
  \cline{2-5}
  & $I_{edge}$ (from CPR measurement) & 82 & 0.0108 &  \\
  \hline
  \multirow{3}{*}{D1 (2023-09)} & $I_{bulk}$ (from T measurement) & 82.1 & 0.00270 & 51 \\
  \cline{2-5}
  & $I_{bulk}$ (from CPR measurement) & 72 & 0.0024 & 45 \\
  \cline{2-5}
  & $I_{edge}$ (from CPR measurement) & 24 & 0.0124 &  \\
  \hline
  \multirow{3}{*}{D2 (2023-09)} & $I_{bulk}$ (from T measurement) & 83.8 & 0.00275 & 52 \\
  \cline{2-5}
  & $I_{bulk}$ (from CPR measurement) & 79 & 0.0026 & 49 \\
  \cline{2-5}
  & $I_{edge}$ (from CPR measurement) & 17 & 0.0089 &  \\
  \hline
\end{tabular}
\caption{Thouless energy $E_{\rm{Th}}$, diffusion constant $D$ and ratio $\frac{10.82E_{\rm{Th}}}{eR_NI_C}$ where relevant, for all SQUID junctions. }
\label{table:allSQUIDs}
\end{table*}

\begin{table*}
\begin{tabular}{|c|c|c|c|c|c|c|}
  \hline
  \multirow{2}{*}{SQUID} & \multirow{2}{*}{Thickness (nm)} & \multicolumn{2}{|c|}{Length $\times$ Width ($\mathrm{\mu m \times \mu m}$)} & \multicolumn{2}{|c|}{$\Phi_0/S_{bulk}$(G)} & \multirow{2}{*}{$R_NI_c$ ($\mu$V)} \\
  \cline{3-6}
  & & Edge junction & Bulk junction & Expected & Measured & \\
  \hline
  C & 223 & 0.6 $\times$ 0.15 & 0.94 $\times$ 1.46 (see Fig.\ref{fig:transformation_hourglass}) & 8 & 2* & 7 \\
  \hline
  A (2022-10) & 282 & 0.7-1.5 $\times$ 1.3 & 0.6 $\times$ 1 & 19 & 10* & 2 \\
  \hline
  B1 (2023-01) & 108 & 0.95-1.5 $\times$ 0.8 & 0.6 $\times$ 1 & 19 & 17 & 3 \\
  \hline
  B2 (2023-01) & 116 & 1-1.5 $\times$ 0.7 & 0.6 $\times$ 1 & 19 & 17 & 2 \\
  \hline
  D1 (2023-09) & 32 & 2 $\times$ 0.55 & 0.5 $\times$ 1 & 21 & 20 & 1.5 \\
  \hline
  D2 (2023-09) & 30 & 2 $\times$ 0.5 & 0.5 $\times$ 1 & 21 & 20 & 1.5 \\
  \hline
\end{tabular}
\caption{Crystal thickness, length and width of the junctions, expected decay field, computed as $\Phi_0$ divided by the bulk junction area, and $R_N I_c$ product of all SQUIDs. The "measured $\Phi_0/S_{bulk}$" value is the perpendicular field at which the critical current is minimum, corresponding to the first zero of a Fraunhofer pattern. We use this minimum critical current criterion even for SQUIDs B and D, that do not display an exact Fraunhofer dependence,  but still present a close to zero critical current followed by a rebound. We include the magnetic focusing by the superconducting electrodes in the estimate of $S_{bulk}$. For all SQUIDs except C, with $L$ the length of the junction and $W$ its width, $S_{bulk} = L \times W + \frac{W^2}{2}$ \cite{Amet}. For SQUID C, we compute the area the same way but by taking the equivalent rectangular junction described in Appendix C. Values with * are not measured directly but estimated, indeed for SQUID C and A, the measurements were done with a large parallel magnetic field and a small perpendicular contribution. The relevant magnetic field here is the perpendicular component, the ratio between the perpendicular component and the total magnetic field is estimated by comparing the SQUID oscillations period with the expected period given the area defined by the SQUID loop.}
\label{table:AFMSQUIDs}
\end{table*}

\begin{figure}[tb]
    \centering
    \includegraphics[width = \columnwidth]{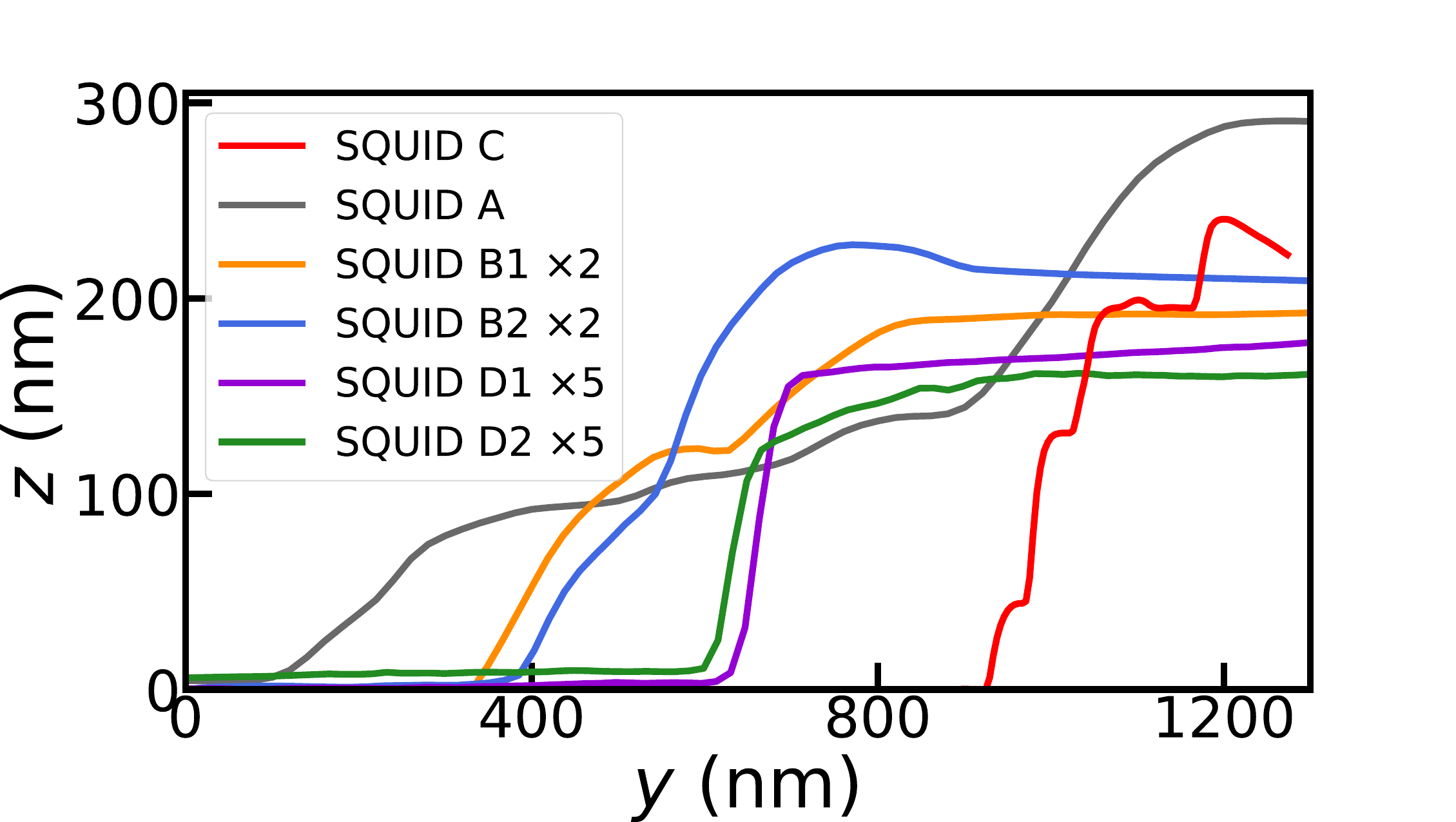}
    \caption{Atomic Force Microscopy measurements of the profiles of all WTe$_2$ crystals of SQUIDs A to D, in the region of the edge junctions. The profile in the perpendicular axis $z$ has been multiplied by two for SQUID B1 and B2, and by 5 for SQUIDs D1 and D2. The thickness of all crystals at the edge junctions are listed in Table \ref{table:AFMSQUIDs}, as well as the length and width of the junctions, the width being defined by the lateral extension of the superconductor. The edge junctions other than SQUID C are spread over a much great lateral area, which means that a large portion of diffusive transport via non topological bulk states is probed by the SQUIDs other than C, providing a possible explanation for the observed sinusoidal CPR.}
    \label{fig:AFMallsteps}
\end{figure}
 
\subsection{Layout and SQUID patterns}
We have fabricated a total of six SQUID samples, which all consist of one junction on the bulk of the WTe$_2$ crystal surface and one overlapping the crystal $a$-edge. The WTe$_2$ crystals have the same origin, and their thicknesses range between 30 and 300 nm, see Table \ref{table:allSQUIDs}. As shown in Fig. \ref{fig:Fig_1}, in contrast to the sawtooth modulation of SQUID C discussed extensively in the main text, two other behaviors are found: Four other SQUIDs display a critical current of the reference junction of several microA and a small, 200 nA modulation, but the modulation is sinusoidal, not sawtooth-like as in SQUID C. Finally, SQUID A displays a symmetric SQUID-like pattern, with a practically full modulation of the critical current. 

\begin{figure}[tb]
    \centering
    \includegraphics[width=0.95\textwidth]{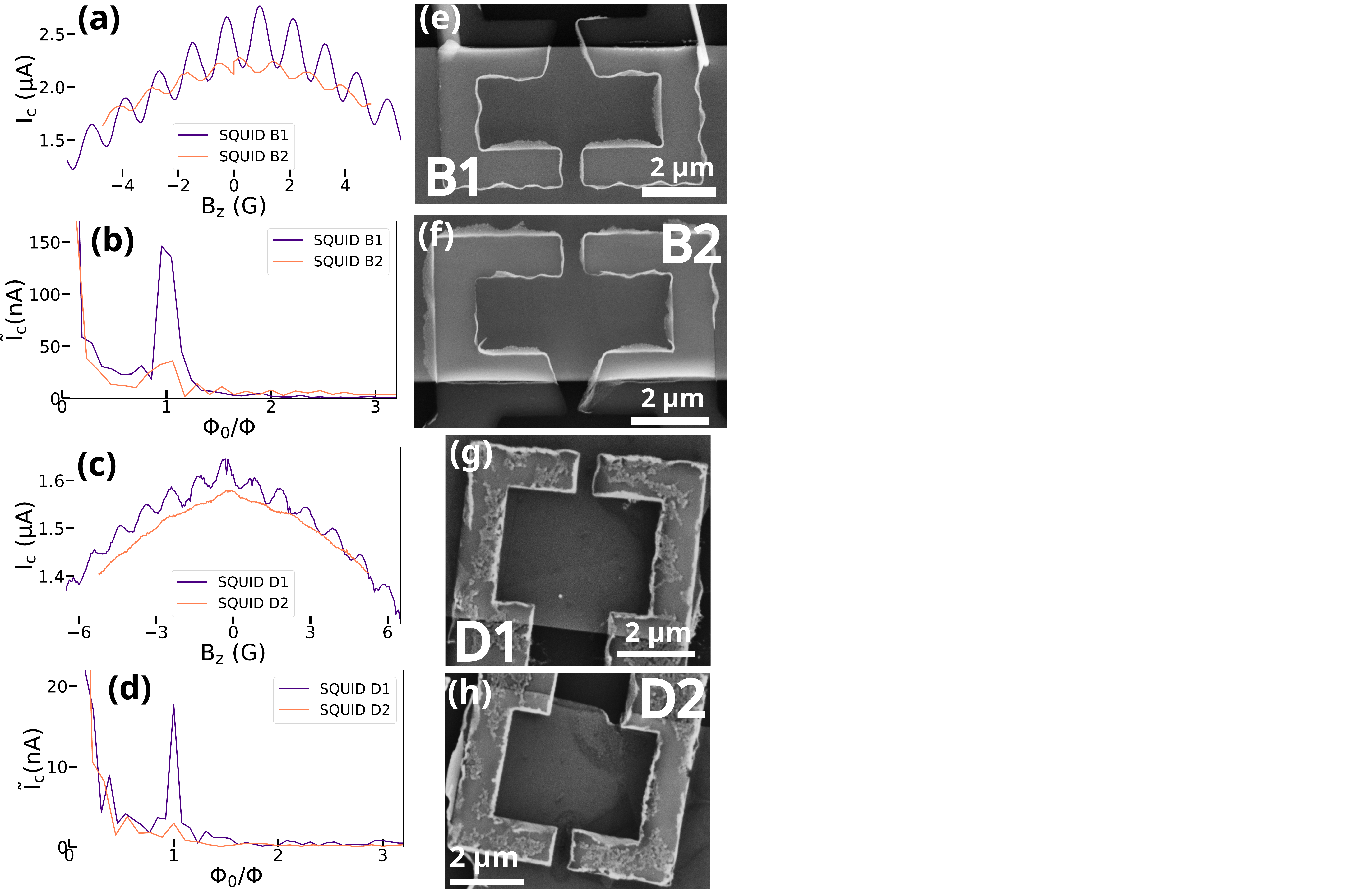}
    \caption{Sinusoidal CPR of SQUIDs B and SQUIDs D's edge junctions. B1 (e) and B2 (f) on the one hand, and D1 (g) and D2 (h) on the other,  are formed on opposite $a$ edges of WTe$_2$ flakes. Like the other SQUIDs shown in \ref{fig:Fig_1}, the superconducting ring is made of a 8 nm-thick layer of palladium covered by a 80 nm-thick layer of niobium. Those four SQUIDs show small sinusoidal oscillations on top of the bigger contribution of the bulk junction. The curves for B1 and B2 (a) are taken at 7 mK while the ones for D1 and D2 (c) are taken at 200 mK. In both cases, the Fourier transform show only one clearly visible peak (b)(d), unlike the high harmonics content of SQUID C's edge junction (Fig. \ref{fig:Fig_2}). Possible explanations for the absence of sawtooth-shaped oscillations (and more harmonics in the FFT) include the greater length of the edge junctions; a different, more dissipative environment; or simply the absence of an electrical contact between the superconductors and the 1D states.}
    \label{fig:SM_other_squids}
\end{figure}

There are several possible reasons that can explain sinusoidal rather than sawtooth behaviors for these SQUIDs.
A first possibility is if in fact no edge states are connected by the superconducting electrodes on the edge junction. Then the SQUID patterns reflect the interference of two bulk junctions in parallel, i.e. the interference of two rather wide diffusive junctions. If the area of both junctions are similar, as in the case of SQUID A (see image in Fig. \ref{fig:Fig_1} and Table II), and if the interface quality is also similar, the SQUID pattern is that of a symmetric SQUID, i.e. $I_c\propto\lvert \cos(\pi \Phi/\Phi_0) \rvert $. The decay of the pattern is governed by the flux through the junction area, on the order of 20 Gauss (taking into account field focusing), as observed in the experiments, see Table II. 
Similarly, the small sinusoidal modulations of SQUIDs B1, B2, D1 and D2 can be explained by an absence of contact to edge states and a weak contact to the diffusive states  at the surface in the junction. 
Figure 26
displays the topography of all SQUID profiles at the edge junctions, and is complemented by Table II, giving the geometry of all junctions. It is clear that the lateral extension of the edge junctions other than that of SQUID C are much greater, meaning that a large portion of diffusive transport via non topological bulk states is probed by the SQUIDs, leading to the observed sinusoidal CPR. 
Table II also shows that SQUID C's edge junction is the shortest of all, so that in fact the 1D hinge states may not be robust and phase coherent over much more than 600 nm, explaining why a sizable supercurrent cannot travel through edge junctions that are longer than 600 nm. 
In addition, a dissipative environment may round a sawtooth shaped supercurrent that is too small, transforming it into a sinusoidal shaped CPR. These environments may depend on the aspect ration of the SQUID loops. The loop of  SQUID C is $1.2 \cross 12 ~\mu m^2$, whereas the loop of SQUIDs B1 (and B2, D1 and D2) have a $2 \cross 5 ~\mu m^2$ area. Since the SQUID loops are not empty but are filled with WTe$_2$, the two junctions are in fact in parallel with bulk WTe$_2$, whose normal resistance acts as a dissipative environment for the junctions. Such a dissipative environment is known to modify the behavior of Josephson junctions, especially those with small supercurrents \cite{ambegaokar_voltage_1969}, decreasing the supercurrent and rounding the CPR via finite-temperature broadening effects \cite{bardeen1972,DellaRocca2007}, leading to sinusoidal CPRs.

Finally, we also note that the sinusoidal modulations in samples with the same reference and edge supercurrents as the sawtooth-shaped modulation are a useful confirmation that the sawtooth CPR of SQUID C is not attributable to an inductive effect,  as in \cite{endres2022}.

\subsection{Comparison of the temperature dependence of the six SQUIDs' bulk and edge junctions}
Comparing the decay with temperature of the critical current of the bulk and edge junctions of all measured SQUIDs (see Fig. 28) 
reveals that only the edge junction displaying a sawtooth behavior is more robust (i.e. has a greater Thouless energy) than the bulk junction. The respective identification of bulk and edge junctions in a given SQUID is possible thanks to the interference pattern in a magnetic field: At low field, the SQUID's critical current is the sum of the bulk junction's critical current and a smaller modulation due to the edge junction's CPR, whose amplitude is the edge junction's critical current. Fourier transforms of those relations display a large zero-frequency peak and a smaller finite-frequency peak related to the small oscillations with a period given by the SQUID loop area. Those peaks yield the amplitudes of both contributions, which we plot as a function of temperature in Fig. \ref{fig:Fig_suppmat_IcT}. Except for the edge junction of SQUID C, The critical current versus temperature curves are then fitted with the dependence expected for a long diffusive junction, yielding the Thouless energy for both edge and bulk junctions, given in Table \ref{table:allSQUIDs}. To this end we fit to one of those two expressions \cite{PhDDubos}
\begin{equation}
    \begin{split}
       eR_NI_c(T)= & \frac{32}{3+2\sqrt{2}} E_{\rm{Th}}(\frac{2\pi k_B T}{E_{\rm{Th}}})^{\frac{3}{2}} \\
        &\times \sum_{n=0}^{n=\infty} \sqrt{2n+1}\mathrm{e}^{-\sqrt{2n+1}\sqrt{\frac{2\pi k_B T}{E_{\rm{Th}}}}}
    \end{split}
\end{equation} 
valid for a long diffusive junction at high temperature $T>5E_{\rm{Th}}$ or

\begin{equation}
   eR_NI_c(T)=10.82 E_{\rm{Th}} (1-1.3\mathrm{e}^{-\frac{10.82 E_{\rm{Th}}}{3.2k_BT}})
\end{equation} 
valid for a long diffusive junction at low temperature $T<5E_{\rm{Th}}$

The two expressions give almost the same $E_{\rm{Th}}$ for all the SQUIDs, leading us to use both indiscriminately even though the fitted curves can have aberrant behavior outside of the temperature range of validity.

A second determination of the bulk junction's critical current was also used for SQUIDs D1 and D2 (diamond marks) and yielded similar values of $E_{Th}^{bulk}$. It is based not on the Fourier transforms but on the total  critical current measured at zero magnetic field for different temperatures. Although this critical current is the sum of the bulk and edge junction's critical currents, since the edge junction's critical current is ten to fifty times smaller than the bulk junction's critical current, it is a good approximation of the bulk junction's critical current. 
As seen in the figure, only the current from SQUID C's edge junction, described in detail in the main text, shows a much slower decay of the critical current with temperature.

\begin{figure*}[tb]
    \centering
    \includegraphics[width=0.95\textwidth]{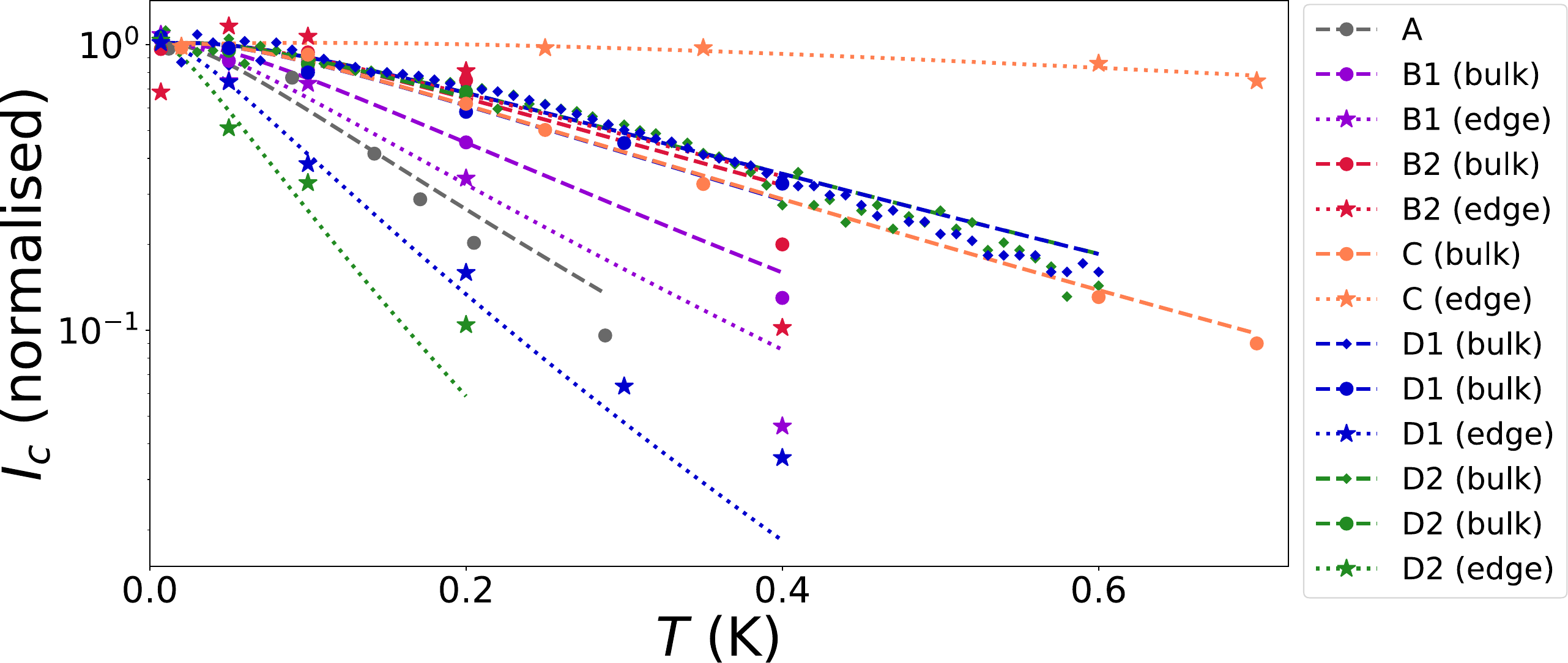}
    \caption{Temperature dependence of the critical current of all measured SQUID junctions. The bulk and edge junctions contributions are identified thanks to the interference patterns of each SQUID and its Fourier transform (circle and star marks). Diamond marks for D1 and D2 correspond to the the total critical current measured at zero field, that is practically equal (within a few percent) to the bulk junction's critical current. The critical current versus temperature curves are then fitted with the dependence expected for a long junction, yielding the Thouless energy for both edge and bulk junctions, given in Table 1. Only the current from the edge junction of SQUID C, described in details in the main text, displays a much slower decay than the bulk junction. The $E_{\rm{Th}}$ are obtained by assuming a very long diffusive junction, see corresponding paragraph in these Supplementary Materials.}
    \label{fig:SM_all_T_variation}
    \end{figure*}

\subsection{Estimate of $WTe_2$'s bulk resistivity}
The two squids B1/B2 and the two squids D1/D2 are respectively both on the same flakes B and D of $\mathrm{WTe_2}$. The bulk junction between the two squids presents a dissipationless current vanishing quickly in field and temperature. By measuring the normal resistance $R_N$, the length $L$, the width $W$ and the thickness $d$ of the junction we can estimate the resistivity $\rho$.

\begin{equation*}
    \rho_B=\frac{Wd}{L}R_N=\frac{1.5\mathrm{\mu m}\times105\mathrm{nm}}{2\mathrm{\mu m}}0.8\mathrm{\Omega} = 6.3\times10^{-8}\mathrm{\Omega.m}
\end{equation*}
\begin{equation*}
    \rho_D=\frac{Wd}{L}R_N=\frac{2.1\mathrm{\mu m}\times35\mathrm{nm}}{1.2\mathrm{\mu m}}1.8\mathrm{\Omega} = 1.1\times10^{-7}\mathrm{\Omega.m}
\end{equation*}
These values agree well with those reported in \cite{choi2020}.

\section{Appendix E: Estimate of inductance effects}
Here we argue that it is unlikely that the sawtooth-shaped CPR we observe in SQUID C's edge junction could be a sinusoidal-like CPR that is artifically tilted by inductance effects. Indeed, inductances (both geometric $L_{g}$ and kinetic $L_{K}$) are known to induce a phase drop proportional to the current $L_{g,K}I/\Phi_0$ in superconductors with a current $I$ running through them, so that the difference in phase drop across the two Josephson junctions in a dc SQUID can differ from the value estimated via the external flux $\Phi_{ext}$ (see Fig. 29 
for the sketch and notations):
$\varphi_1-\varphi_2=2\pi\Phi_{ext}/\Phi_0-(L_1I_1-L_2I_2)/\Phi_0$.
If the inductances and currents are large enough, and/or the asymmetry of inductance and/or current between the two junctions  is large enough, this can lead to tilted current-versus-external flux relations $i(\Phi_{ext})$ that can be misinterpreted as tilted current versus phase relations $i(\varphi)$ \cite{Tinkham,endres2023,alexandrebernard2022}. These effects become important when $L_{g,K}i/\Phi_0\gg1$.

\begin{figure}[tb]
    \centering
    \includegraphics[width=0.65\columnwidth]{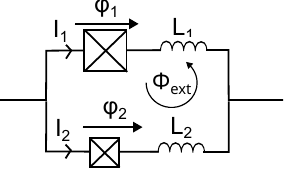}
    \caption{Schematics of a SQUID with an inductance in each arm. The upper arm, labeled 1, contains the reference junction. The other arm, labeled 2, contains the edge junction. }
    \label{fig:SM_inductance_effect}
\end{figure}

\begin{figure*}[tb]
   \centering
    \includegraphics[width=0.9 \textwidth]{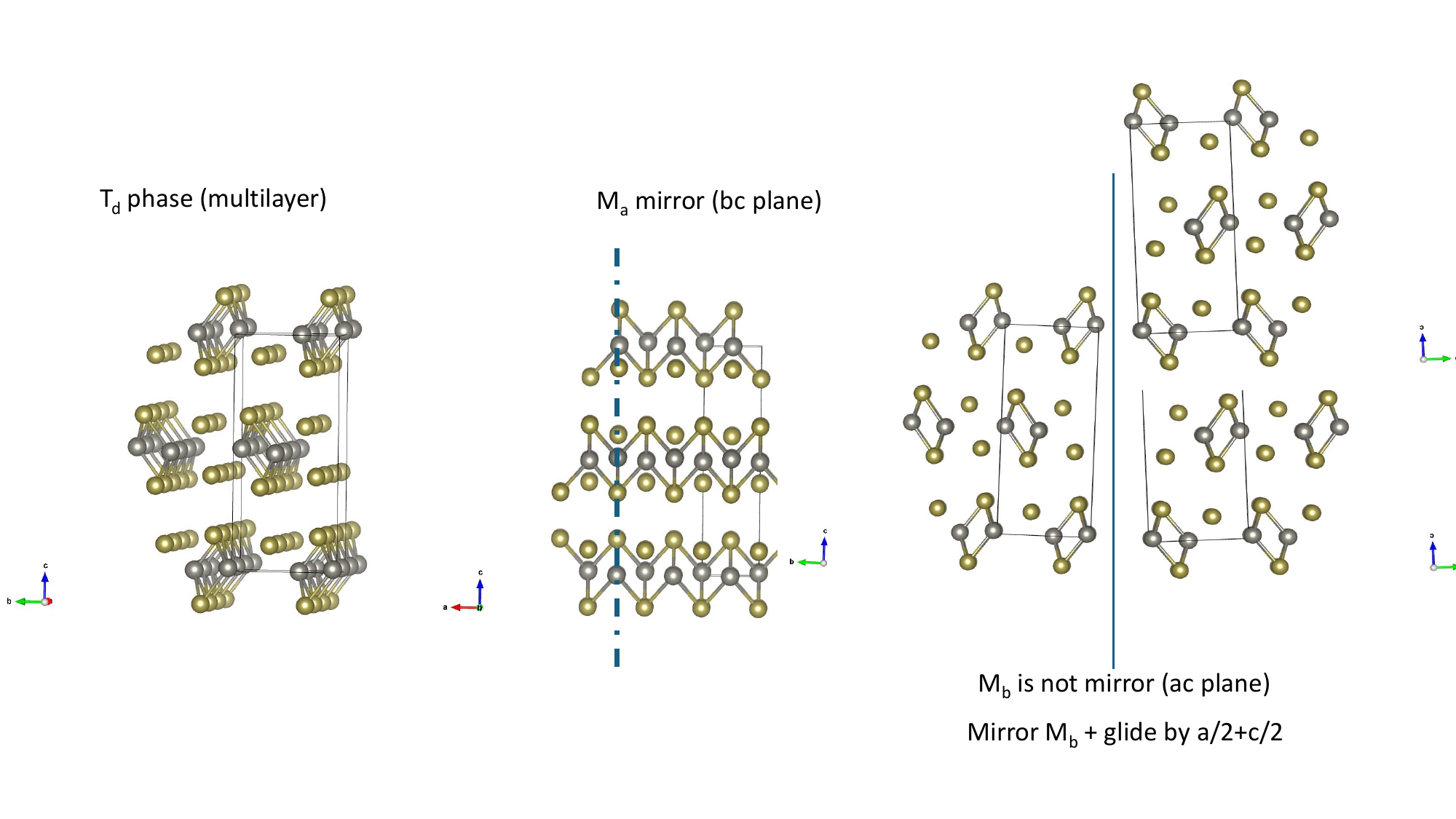}
    \caption{Representation of the multilayer WTe$_2$ atomic arrangement}
    \label{fig:SM_symmetries}
\end{figure*}

An estimate of SQUID C's loop's geometrical inductance is $L_g=\SI{14}{\pico\henry}$ \cite{Shatz2014}, given  the loop dimensions $\SI{18}{\micro\meter}\times \SI{4}{\micro\meter}$ and thickness $\SI{90}{\nano\meter}$. An estimate of the loop's kinetic inductance is   $L_K=\SI{9}{\pico\henry}$ \cite{Annunziata2010}, given the square resistance of the Nb film $R_{\square}\simeq 2\mathrm{\Omega/\square}$.
Assuming both arms are roughly identical yields a total (kinetic and geometric) inductance : $L_1\approx L_2\approx \SI{12}{\pico\henry}$.
At worst, the difference in supercurrent in both arms can be estimated as the maximum critical current of the reference junction: $I_1-I_2\approx \SI{4}{\micro\ampere}$. Then the parameter characterizing the importance of inductance effects is \cite{Tinkham} $\beta_m=\frac{L_1I_1-L_2I_2}{\Phi_0}\approx 0.03\ll1$, implying that inductive effects are far from sufficient to induce a deformation of the CPR such that a sinusoidal CPR would appear as a sawtooth CPR.

In ref. \cite{endres2023} where a similar system is studied, the author find an additional inductance of the order $L\approx\SI{100}{\pico\henry}$, that is attributed to a PdTe compound that has formed at the interface between Pd and WTe2. Even allowing for such a large inductance in the arm containing the edge junction, the parameters would be $L_2= \SI{112}{\pico\henry}$,  $I_2=\SI{150}{\nano\ampere}$, $L_1=\SI{12}{\pico\henry}$ and $I_1=\SI{4}{\micro\ampere}$ , and the importance of screening effects would be quantified by $\beta_m\approx 0.02$, which is negligible. 

In addition, maybe the most convincing argument against a large effect of inductances in our measurements is the fact that we $\it{do}$ measure sinusoidal CPRs in some of the SQUIDS (B1, B2 and D1 and D2), that have similar contacts, geometry and critical currents, and do not always measure sawtooth CPRs. 

\section{Appendix F: Synthesis of WTe$_2$ crystals}

The single crystals of WTe$_2$ were grown through the flux growth technique from a $95\%$ Te-rich melt. High-purity W powder (Alfa Aesar, $99.999 \%$) and Te ingots (Sigma Aldrich, $99.999 \%$) were used as the precursors. The obtained crystals were characterized using X-ray diffraction (XRD) and energy-dispersive X-ray spectroscopy (EDX). To get a qualitative estimate of the crystallographic defect density, we measured the temperature dependence of the longitudinal resistivity of several randomly selected crystals. Within the 1.8 to 300 K temperature range. All WTe$_2$ single crystals show a residual resistivity ratio (RRR) of 1000-1200, implying their high crystalline quality.

\section{Appendix G: Symmetries of WTe$_2$}

$\mathrm{WTe_2}$ is a transition-metal dichalcogenide (TMD) material constituted of layers in the (a,b) plane stacked along the $c$ direction. The atomic arrangement in the (ab) plane is characterised by tungsten atoms forming chains along the $a$ direction, and the absence of a mirror symmetry perpendicular to the b vector\cite{Cava2014}.
Monolayer $\mathrm{WTe_2}$ has been predicted to be a 2D topological insulator in the stable,  inversion-symmetric 1T'-$\mathrm{WTe_2}$ phase \cite{qian2014a,zheng2016}, and this prediction is supported by transport \cite{Fei2016,wu2018a} and local probe \cite{peng_observation_2017,tang2017,shi2019} measurements that report insulating bulk and quantum spin Hall edge conduction.
The stable 3D crystal form of $\mathrm{WTe_2}$ is the $T_d$ phase (symmetry group $Pmn2_1$), which is non-centrosymmetric. It possesses a $M_a$ ($bc$-plane) mirror symmetry, an $\Tilde{M_b}$ ($ac$-plane) glide-mirror symmetry, and a two-fold screw axis along the $c$-axis. The double band inversion at the gamma point combined with the screw symmetry along the $c$-axis gives rise to a bulk 3D topological state for which surface domain walls and hinges can bind helical modes whose configurations are extrinsically controlled by a combination of surface band inversions and disorder  
\cite{wang2019a,song_natcom_2018,wieder_science2018}. This can be understood because double band inversion indicates that the low-energy topological physics is controlled by a massive eightfold (double) 3D Dirac fermion, for which extrinsic mass textures bind 1D helical modes \cite{wieder_double_2016}. Like in BiBr, helical hinge modes in WTe$_2$ along the ab axes likely appear on alternating terraces/step edges, due to a combination of microscopic van der Waals coupling, the 3D bulk double band inversion, and the nontrivial 2D helical (Z2) topology of monolayer $\mathrm{WTe_2}$. \cite{peng_observation_2017}.  However, the band structure of $\mathrm{WTe_2}$ presents conducting electrons and holes pockets with a possible type II Weyl point due to the broken inversion symmetry, which makes the observation of such 1d states in transport experiments challenging.

\end{document}